\documentclass[twocolumn,aps,]{openjournal}

\usepackage{outlines}
\usepackage{enumitem}
\usepackage{comment}
\usepackage{amsthm}
\usepackage{amsmath}
\usepackage{amssymb}
\usepackage{xcolor}
\usepackage{xspace}
\usepackage{microtype}
\usepackage{multirow}
\usepackage{aas_macros}
\usepackage{ulem}
\usepackage{lineno,hyperref}
\modulolinenumbers[5]

\shorttitle{Optical Cluster Cosmology with SBI}
\shortauthors{Reza, Zhang, et al.}

\begin{document}
\label{firstpage}

\def\bn#1{\textcolor{blue}{BN: #1}}
\def\yz#1{\textcolor{red}{YZ: #1}}
\definecolor{cacolor}{RGB}{121,0,102}
\def\ca#1{\textcolor{cacolor}{CA: #1}}
\def\mr#1{\textcolor{green}{MR: #1}}
\def\todo#1{\textcolor{red}{#1}}

\title[Optical Cluster Cosmology with SBI]{Constraining Cosmology with Simulation-based inference and Optical Galaxy Cluster Abundance}


\author{Moonzarin Reza}
\email{moonzarin@tamu.edu}
\affiliation{Department of Physics and Astronomy, Texas A\&M University, College Station, TX 77843, USA\\
George P. and Cynthia Woods Mitchell Institute for Fundamental Physics and Astronomy, Texas A\&M University, College Station, TX 7783, USA}
\author{Yuanyuan Zhang}
\email{yuanyuan.zhang@noirlab.edu}
\affiliation{
Community Science and Data Center/NSF NOIRLab, 950 N. Cherry Ave., Tucson, AZ 85719, USA \\
George P. and Cynthia Woods Mitchell Institute for Fundamental Physics and Astronomy, Texas A\&M University, College Station, TX 7783, USA}
\author{Camille Avestruz}
\affiliation{Department of Physics, University of Michigan, Ann Arbor, MI 48109, USA\\
Leinweber Center of Theoretical Physics, University of Michigan, Ann Arbor, MI 48109, USA}
\author{Louis E. Strigari}
\affiliation{Department of Physics and Astronomy, Texas A\&M University, College Station, TX 77843, USA\\
George P. and Cynthia Woods Mitchell Institute for Fundamental Physics and Astronomy, Texas A\&M University, College Station, TX 7783, USA}
\author{Simone Shevchuk}
\affiliation{Fermi National Accelerator Laboratory, Batavia, IL 60510}
\author{Francisco Villaescusa-Navarro}
\affiliation{Center for Computational Astrophysics, Flatiron Institute, New York, NY, 10010, USA}

\begin{abstract}

We test the robustness of simulation-based inference (SBI) in the context of cosmological parameter estimation from galaxy cluster counts and masses in simulated optical datasets.  We construct ``simulations'' using analytical models for the galaxy cluster halo mass function (HMF) and for the observed richness (number of observed member galaxies) to train and test the SBI method.  We compare the SBI parameter posterior samples to those from an MCMC analysis that uses the same analytical models to construct predictions of the observed data vector.  
The two methods exhibit comparable performance, with reliable constraints derived for the primary cosmological parameters, ($\Omega_m$ and $\sigma_8$), and richness-mass relation parameters. We also perform out-of-domain tests with observables constructed from galaxy cluster-sized halos in the Quijote simulations. Again, the SBI and MCMC results have comparable posteriors, with similar uncertainties and biases. Unsurprisingly, upon evaluating the SBI method on thousands of simulated data vectors that span the parameter space, SBI exhibits worsened posterior calibration metrics in the out-of-domain application.  We note that such calibration tests with MCMC is less computationally feasible and highlight the potential use of SBI to stress-test limitations of analytical models, such as in the use for constructing models for inference with MCMC.

\end{abstract}




\section{Introduction}
\label{sec:introduction}

The large-scale structure of matter in our Universe is a sensitive probe of cosmological models \citep{weinberg2013observational, abbott2022dark}.  
Galaxy clusters provide a tracer of structure that can be used in cosmology analyses \citep{frieman2008dark, 2011ARA&A..49..409A, 2012ARA&A..50..353K, weinberg2013observational}. In particular, the time evolution of the galaxy cluster mass function is sensitive to $\Omega_m$, the energy density of matter in our universe, and $\sigma_8$, the amplitude of the power spectrum on the scale of 8~$h^{-1}\mathrm{Mpc}$. 
Through measuring the abundance and mass distribution of galaxy clusters, cosmic observations have yielded increasingly precise and accurate cosmological constraints \citep[e.g., ][]{2009ApJ...692.1060V, 2010ApJ...708..645R, 2014MNRAS.440.2077M, 2020PhRvD.102b3509A, 2021PhRvL.126n1301T, 2023arXiv230913025S, 2024arXiv240102075B}.

Optical observations of galaxy clusters are an important component of the cluster cosmology endeavor. Large optical surveys, such as the Dark Energy Survey (DES) \citep{abbott2022dark, desai2021dark}, Kilo-Degree Survey (KiDS) \citep{asgari2021kids}, and the Legacy Survey of Space and Time (LSST) \citep{mandelbaum2018lsst} at the Vera C. Rubin Observatory are rapidly increasing the number of identified galaxy clusters that the cosmology analysis with them will soon enter an era that is not statistics-limited. 
On the other hand, discrepancies between cluster-based cosmology constraints and other methods indicate a pressing need to disentangle systematic biases \citep[due to observational limitations or assumptions in modeling as studied and discussed in,  e.g., ][]{2022MNRAS.515.4471W, 2022MNRAS.511L..30Z, 2023MNRAS.521.5064S} from signatures of a $\Lambda$CDM cosmological model and beyond. In particular, traditional methods implemented with the Markov Chain Monte Carlo (MCMC) parameter sampling \citep[e.g., ][]{2020PhRvD.102b3509A, 2021PhRvL.126n1301T, 2023arXiv230913025S} often require an explicit observable likelihood calculation, performed during the parameter sampling process. These methods can become computationally inefficient due to more complicated models to account for various systematic effects. 
The need for more efficient computing methods can thus benefit from the development of methods that use implicit likelihoods. 

Simulation-based inference (SBI) \citep{doi:10.1073/pnas.1912789117} is one such approach.  Here, the method forward models the observables assuming prior distributions on model parameters in the ``simulation'' step.  We can then train a machine learning model on the ``simulations'' to derive the joint probability of model parameters and the observed data vector. 
 We focus on an approach in optical cluster cosmology, in which we model the cluster observables ($x$) based on a $\mathrm{\Lambda CDM}$ cosmology model and a cluster richness \citep[the observed count of red sequence cluster galaxies, ][]{2012ApJ...746..178R, 2014ApJ...785..104R} model with their associated parameters ($\theta$). 
With the SBI approach, we fit a function $P(x, \theta)$ over our simulations $\{\theta, x\}$ resulting in a joint probability model of observables and model parameters.
From the trained probability model, we can obtain the posterior distribution of the model parameters (cosmological parameters and richness model parameters), $P(\theta|x)$, for a given ``observation" of a data vector, $x$. 

SBI has been applied in several domains, such as health analysis \citep{contoyannis2004simulation} and introductory statistics \citep{rossman2014using}. 
In particle physics, SBI has been used to obtain posteriors for measurements of the Higgs boson and in searches for direct signals of heavy new particles \citep{brehmer2021simulation, brehmer2022simulation}. 
In astronomy, SBI has been used to study the cosmic microwave background \citep{lemos2022robust}, to characterize the excess gamma-ray at the Center of the Milky Way 
\citep{mishra2022neural},
to constrain the mass of warm dark matter based on the observed stellar density along the GD-1 stream \citep{hermans2021towards}, and to recover posteriors for the reionization parameters from 3D Tomographic 21 cm images \citep{zhao2022simulation}.

\cite{kodi2021simulation} used SBI to obtain dynamical masses of galaxy clusters found in the Sloan Digital Sky Survey Legacy Survey, yielding results consistent with other mass measurements in the literature. In cosmology, \cite{2022ApJ...925..145T} developed an SBI approach to model galaxy cluster abundance and their weak lensing signals to derive cosmological constraints. Similar methods have also been applied to the adjacent large-scale structure cosmology sub-field: \cite{lin2023simulation} applied an SBI approach to model large-scale structure weak gravitational lensing data from the Kilo-Degree Survey while \cite{2021MNRAS.501..954J, 2024arXiv240302314J} developed the SBI approach to analyze lensing maps from the Dark Energy Survey.


In this study, we apply SBI to optical galaxy cluster observables and constrain five fundamental cosmological parameters and three cluster richness model parameters.
 We aim to test the statistical fidelity of SBI results and identify potential caveats of the SBI method so that future optical cluster cosmology analysis with real-Universe measurements can be prepared to tackle these caveats. This work is based on the precursor analysis by \cite{reza2022estimating} and incorporates the galaxy cluster richness quantity for cluster selection, as well as improved statistical tests. The rest of this paper is organized as follows. In Section~\ref{sec:methods:toymodel}, we describe our simplified model and methodology, starting with analytical cluster observable models to simulate our observable data vector and an introduction to the Quijote simulations \citep{villaescusa2020quijote}, which we will apply our methods to.  In Section~\ref{sec:methods}, we describe SBI and MCMC, the methods we use for parameter inference. In Sections~\ref{sec:results} and~\ref{sec:mcmc}, we discuss the results of our parameter inference and compare the MCMC and SBI methods.
In Section~\ref{sec:result_quijote}, we describe the results of applying the MCMC and SBI methods to test cases constructed with Quijote simulations. Finally, Section~\ref{sec:conclusiondiscussions} summarizes our findings.

\section{Models of Optical Cluster Observables}
\label{sec:methods:toymodel}

In this section, we introduce our models for simulating optical galaxy cluster observables for a cosmological analysis. 
To enable testing on a set of Quijote simulations (described in Section~\ref{sec:sim_quijote}), we choose specifics of the models to match these of the Quijote simulations. 
This also simplifies our analysis to not include many observational effects, such as the universe's volume, the cluster redshift uncertainties, and many of the systematic effects affecting cluster mass measurements, like large-scale structure projections and cluster triaxiality. We leave these detailed modeling for future work and focus on studying the statistical fidelity of SBI in an optical galaxy cluster cosmology analysis. We construct two sets of galaxy cluster simulations: one based on an analytical model of the cluster halo mass function and another based on the Quijote Latin Hypercube simulations. The rest of this section describes how we generate these simulations. 

\subsection{Analytical models for Galaxy Cluster Abundance} \label{sec:analytical_richness_models}

Using analytical models, we generate simulations of optical galaxy cluster observables to use in cosmological analysis. 
Here, we will provide an overview of the models we use to analytically simulate halo mass distributions, and describe how we generate the cluster richness quantities on top of the halo mass simulations. 

First, we generate a halo mass function (HMF) for a set of cosmology parameters and draw samples of halos (We use the terms, ``halo'' and ``galaxy cluster,'' inter-changeably) from the HMF with a given mass range and redshift.
The halo mass function (HMF) describes the number density $n(M, z)$ of halos with mass $M$ and redshift $z$. 
We use the \cite{2011ApJ...732..122B} HMF for halo's friends-of-friends mass definition \citep{1984MNRAS.206..529E} in the Colossus package \citep{2018ApJS..239...35D}. When examining halo counts in the simulations, we found that the Bhattacharya function, among the HMFs in Colossus, closely matches the halo mass distribution of the Quijote Latin Hypercube simulations, which we will also explore in this paper.  We create two halo samples, one at each redshift, $z=0$ and $z=0.5$, with sample sizes corresponding to the expected number of halos in a cosmic volume of $1\mathrm{Gpc}^3 h^{-3}$, which is the volume of the Quijote simulations.

Next, we assign a richness, $\lambda$, to each dark matter halo. This quantity refers to a probabilistic number count of red sequence galaxies found in a galaxy cluster and can be used as a mass proxy in optical cluster cosmology studies \citep{rykoff2012,murata2019}.  We assume a log-normal probability distribution function for the cluster richness-mass relation $P(\lambda|M)$, such that $\mathrm{ln}{\lambda} \sim \mathcal{N}(\overline{\mathrm{ln} {\lambda}}, \mathrm{ln}\sigma)$. The mean of the distribution depends on the halo mass: $ \overline{\mathrm{ln} {\lambda}} = M_A + M_B \mathrm{ln} (M)$, where $M_A$ is a constant and $M_B$ scales the richness dependence on the halo mass $\mathrm{ln} (M)$. The standard deviation of the normal distribution, $\mathrm{ln} \sigma$,  adds scatter to the cluster richness quantities. We adopt these relations based on the analysis in \cite{baxter2016constraining}, which constrained cluster richness-mass relations using 8312 redMaPPer clusters \citep{2014ApJ...785..104R} identified in the Sloan Digital Sky Survey (SDSS) data \citep{2011ApJS..193...29A, 2011ApJS..195...26A}.

With the analytically simulated halo samples, we have the flexibility of choosing the halo mass function and the redshift values. We can also quickly generate cluster simulations to test SBI and MCMC results at arbitrary points in the parameter space. The MCMC method implemented in this paper uses the same model ingredients as these simulations (See description in Section~\ref{sec:mcmc_method}), enabling a consistent comparison between MCMC and SBI methods in this paper.

\subsection{Quijote simulations} 
\label{sec:sim_quijote}

The Quijote simulations \citep{villaescusa2020quijote} are a suite of 44,100 full N-body simulations developed for cosmological studies. The simulations use 256$^3$ , 512$^3$ , or 1024$^3$ particles to simulate the growth of cosmic structures in a volume of 1 h$^{-3}$~Gpc$^3$. 
This data set is especially useful for training machine learning methods to use for analyzing large-scale structures in the universe.

The Quijote simulation suite contains subsets of simulations that were generated with different cosmological models, initial conditions, and simulation resolutions. 
We use the Latin Hypercube simulations, which contain 2000 universes simulated with a $\Lambda$CDM cosmology model of varied cosmological parameters drawn from uniform distributions (same with the prior distribution listed in Table~\ref{tbl:AnaTests}), with a resolution of 512$^3$ particles within a volume of $1 (\mathrm{Gpc}h^{-1})^3$.

To simulate galaxy cluster observables, we use dark matter halos identified by the friends-of-friends (FoF) algorithm at $z=0$ and 0.5 with a minimum mass of $1\times10^{14}$ h$^{-1}$ \(M_\odot\). 
The mass threshold ensures that each halo will contain at least 20 dark matter particles regardless of the varying particle mass resolution in the simulations. For consistency, we apply the same halo mass threshold to the analytical models (Section~\ref{sec:analytical_richness_models}) and to the MCMC methods. Similar to the analytical simulations, we then compute a \textit{richness} quantity to each of the dark matter halos, using an analytical richness-mass relation. This procedure is the same with the application on the analytically simulated halo masses, as described in Section~\ref{sec:analytical_richness_models}. With these steps,  we also derive samples of galaxy clusters from the Quijote Latin Hypercube simulations, each with a mass and richness quantity.

\subsection{Observable Data Vector in Galaxy Cluster Richness Bins} \label{sec:binned_richness}

From the cluster mass and richness simulations, we derive summary statistics that will comprise the observable data vector for our cosmological analysis.
Our data vector consists of the counts of and the average masses of galaxy clusters in richness bins at different redshifts. 
This binning approach is designed according to galaxy cluster cosmology analyses already applied to the Sloan Digital Sky Survey\citep{2010ApJ...708..645R, 2019MNRAS.488.4779C}  and the Dark Energy Survey \citep{2020PhRvD.102b3509A}, that measure counts and average masses \citep[usually through weak-lensing][]{2009ApJ...703.2217S, 2007arXiv0709.1159J, 2017MNRAS.466.3103S, 2019MNRAS.482.1352M} of galaxy clusters in richness bins, and then derive cosmology parameters by forward modeling these observables from theoretical functions such as the halo mass function, and richness-mass relations, etc. 

To mimic the richness-binning approach, we take the $z=0$ clusters of one simulation, and find clusters with richness values in the following ranges: $[20, 45]$, $[45, 80]$, $[80, 200]$ and $[200, +\infty]$. For each richness range, we count the number of halos in that range and derive their average halo masses. We repeat the binning procedures for the redshift 0.5 clusters of the same simulation, as richness-based galaxy cluster cosmology analyses often study a broad redshift range under redshift 1.0. Thus, for each simulation, we derive a 16-element observable data vector, corresponding to the total number counts and average masses of the clusters in four richness bins at two redshifts. We also refer to this observable data vector as a summary statistics data vector, as it has compressed observations of individual clusters into a smaller, informative data vector. These data vectors are analyzed by both the SBI and MCMC methods to derive cosmology and cluster richness-mass relation parameters.

\begin{figure*}
\includegraphics[width=1.0\columnwidth]{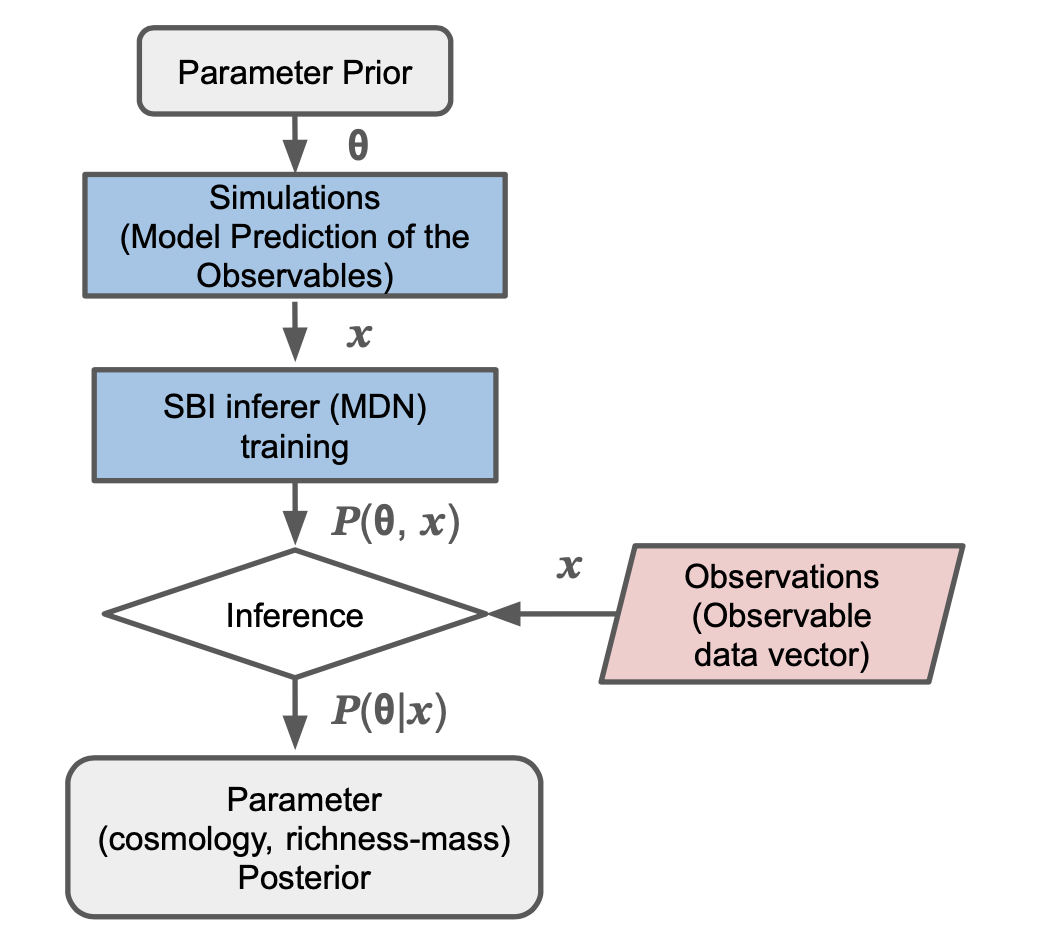}
\includegraphics[width=1.0\columnwidth]{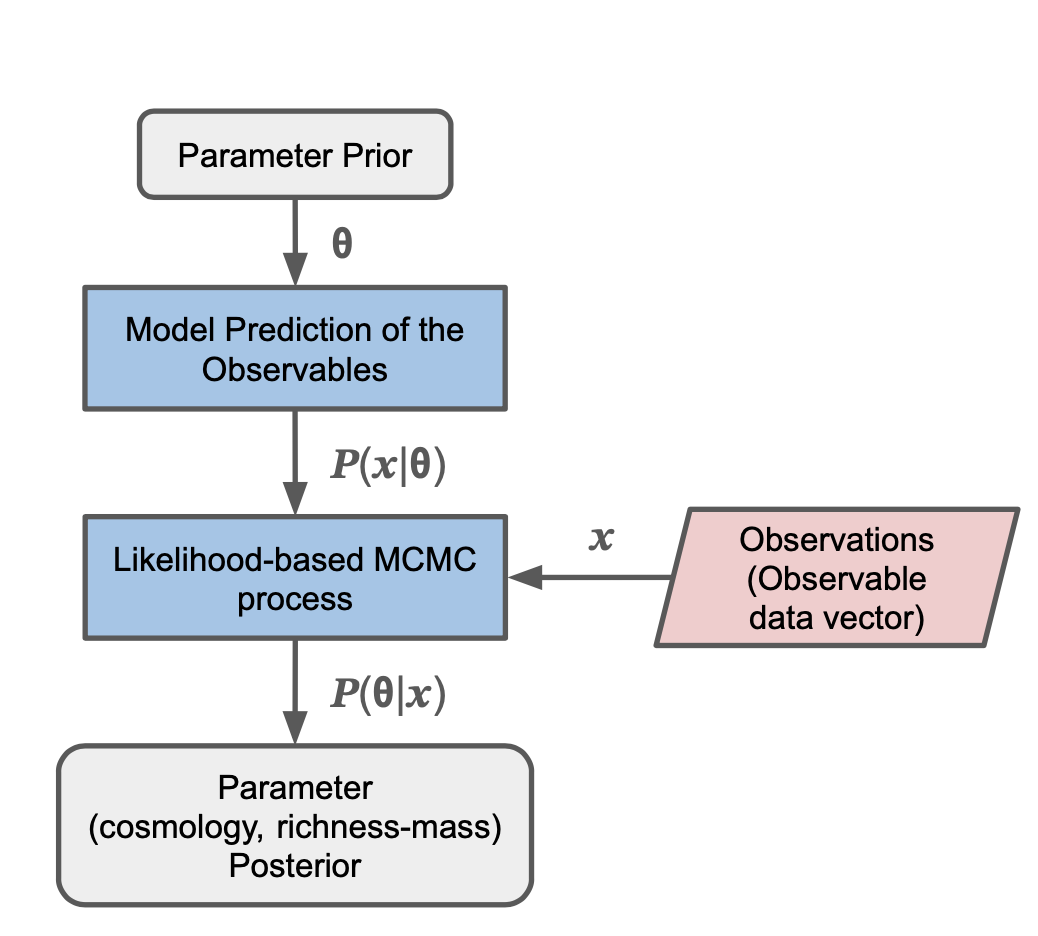}
    \caption{
    Left: Flowchart illustrating the SBI analysis procedure. In the figure, $\theta$ refers to the model parameters and $x$ refers to the observable data vector. We sample model parameter values from their prior ranges to generate simulations of the observable data vector. We then use a mixture density network (MDN) to learn the joint probability function between the parameters and the observables. The trained MDN is used to generate a posterior sample of the model parameters for a given set of observable data vector. Right: Flowchart illustrating the MCMC process. The MCMC process depends on assessing the likelihood of the observables for a given set of parameter values (proposed from the parameter prior ranges) to generate a posterior parameter sample.
    }
    \label{fig:sbi_block}
\end{figure*}

\section{Methods for Analyses}
\label{sec:methods}

We test the performance of SBI on applications to galaxy cluster observables described in Section~\ref{sec:methods:toymodel}.
To better understand the statistical fidelity of SBI, we compare its performance to that of a more traditional method, Markov Chain Monte Carlo (MCMC). 
This section describes the details about the setup of the SBI and MCMC methods. 

\subsection{SBI Simulation Generation}
We use analytical models (Section~\ref{sec:analytical_richness_models}) to generate 200,000 universes, each with parameters, $\Omega_m$, $\Omega_b$, $h$, $n_s$, $\sigma_8$, $M_A$, $M_B$, and $\mathrm{ln}\sigma$ randomly drawn from a flat prior distribution (Table~\ref{tbl:AnaTests}).
The ranges of the five cosmological parameters match the  parameter ranges in the Quijote Latin Hypercube simulation suite. 
The three richness-mass parameters are selected based on \cite{baxter2016constraining}.

We generate the simulations with the analytical models using computing resources from the Open Science Grid (OSG) \footnote{https://osg-htc.org}, which is a distributed computing consortuim with publicly available resources. Because each analytical simulation can be independently generated, our application can efficiently make use of the OSG high-throughput computing resources in an opportunistic mode. 
We have used 1000 parallel processes to generate 200,000 sets of simulations in fewer than three days. 

We also use the Quijote simulations to generate observable data vectors for testing and training purposes. The Quijote Latin Hypercube simulation suite contains 2000 N-body simulations, each with unique cosmological parameters. To generate sufficiently large training and testing sets, for each of the Latin Hypercube simulations, we run 100 iterations of richness generation, varying the values of $M_A$, $M_B$, and $\mathrm{ln}\sigma$ in each iteration. With these iterations, we acquire 200,000 sets of simulated universes  to use for SBI training and testing. 

\subsection{Setup of the Simulation-based Inference Method} \label{sec:sbi_setup}

\begin{figure*}
 \centering
\includegraphics[width=2.2\columnwidth]{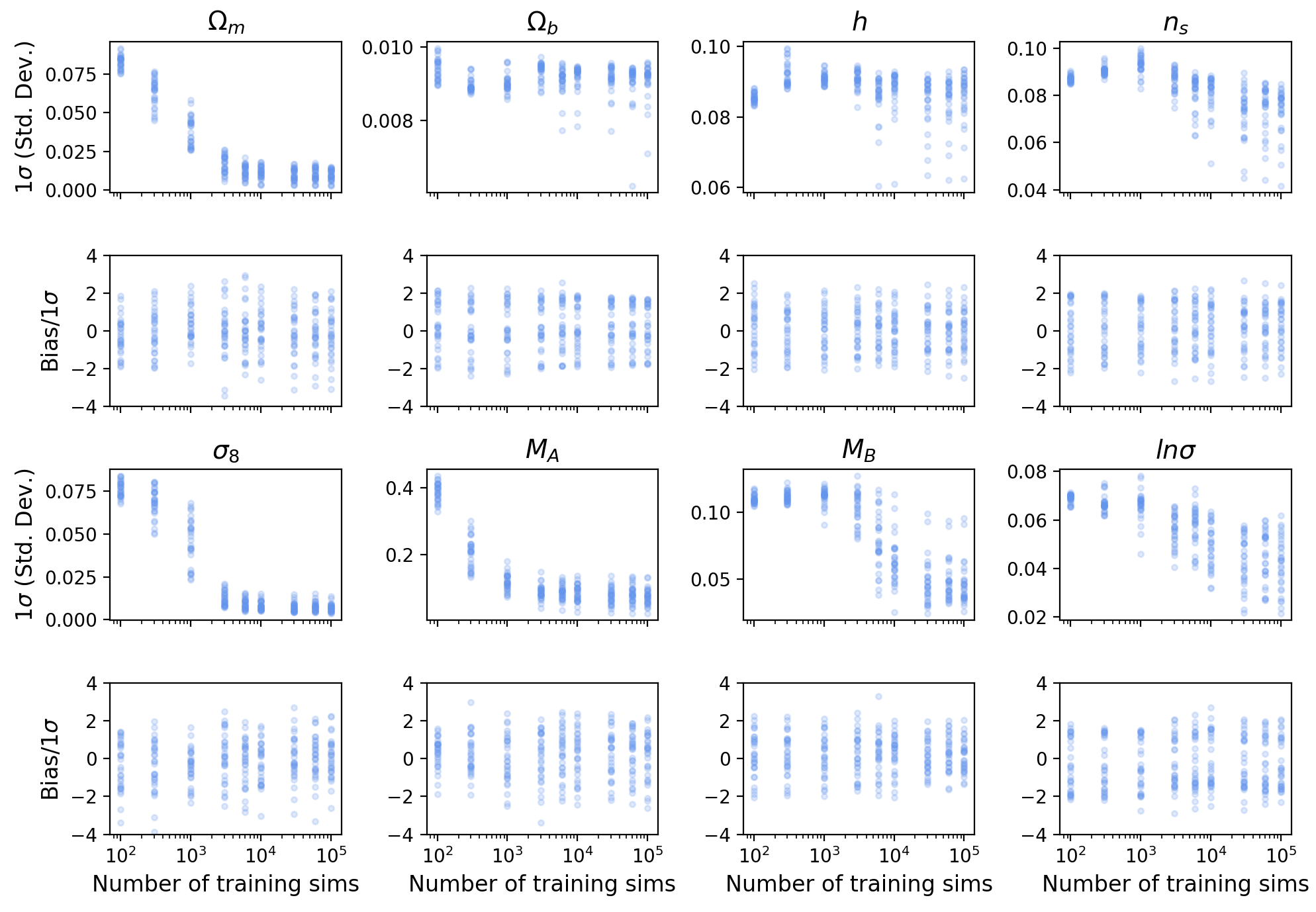}   
    \caption{
    Model performance metrics as a function of the number of simulations used in training.
    {\it First and third Rows}: We note that the parameter posterior uncertainties from our SBI application depend  on the size of the training set. In our application, the parameter posterior uncertainties derived from the SBI method reduce and then stabilize as an increasingly large number of simulations is used to train the SBI method. After the training  size reaches 10,000 sets of simulations, we no longer notice an obvious uncertainty reduction. We interpret the trend as that a training set larger than 10,000 simulations likely contains sufficient information for our application to learn the probability distribution between parameters and observables. {\it Second and Fourth row}: The parameter posterior biases from the SBI method appear to be independent of the size of the training set. For our SBI application, we use a total number of 150,000 sets of simulations for training. See Section~\ref{sec:sbi_setup} for more detailed discussions. 
    }
    \label{fig:Unc_numSims}
\end{figure*}

The SBI application in this paper is implemented with the SBI package \citep{tejero2020sbi}, a neural network-based, publicly available PyTorch \citep{ketkar2021introduction}  package\footnote{https://github.com/sbi-dev/sbi}. The process of the application is broadly illustrated in Figure~\ref{fig:sbi_block}.

The simulations are divided into a training set containing 150,000 sets of simulations, and a testing set containing 50,000 sets of simulations. 
Unless otherwise specified, the default SBI method used in this paper is trained and tested with the analytical simulations, but we also tested the SBI method trained with Quijote-based simulations. 
The training set of simulations, with their simulated observables and model parameters, are used to train a mixture density network (MDN), which is a machine-learning network that determines a probabilistic relationship between the parameters and the observables. 
We also tested some of the analyses using an alternative Masked Autoregressive Flow (MAF) algorithm. See the Appendix for results. 
We denote the model parameters as $\theta$, and the observables as $x$. 
The SBI training process fits an MDN model that represents the joint distribution of $\theta$ and $x$ as $P(x, \theta)$. 
After acquiring such an algorithm (the trained MDN), upon a given set of observables, the algorithm can be directly used to sample the model parameters's posterior probability function $P(\theta|x)$. 
We use the default posterior estimation method (SNPE) in the SBI package for these procedures, and derive 10,000 posterior samples of the model parameters, to represent the parameter's posterior distribution. 

Previous applications of the SBI process \citep{reza2022estimating, 2024arXiv240302314J} has noted that the parameter posterior sample distributions can be sensitive to the number of training simulations, especially when the number of training simulations is small. SBI algorithms trained with a small set of training simulations can produce a broad range of posterior parameter values, broader than expected from the Bayesian probability functions. This is likely due to insufficient information for the SBI inference algorithm to fully learn the joint distribution of the parameters and observables, $P(x, \theta)$. We examine how the width of the parameter posterior distribution changes with the number of training simulations, to determine whether or not the MDN has been trained with a sufficiently large training set. 

We use training sets with the following sizes $N_{train} \in [100, 300, 1000, 3000, 6000, 10000, 30000, 60000, 100000]$.
For each experiment, we retrain the SBI model and study the performance on test sets. 
From the parameter posterior samples, we derive the 1$\sigma$ standard deviation of each parameter, as well as a bias quantity, which is the difference between the parameter's posterior mean value and the parameter's truth value in the test case. We examine how these 1$\sigma$ standard deviation and bias quantities change with the number of training simulations.  
We evaluate these quantities for 30 tests randomly drawn from the test set.
The result is shown in Figure \ref{fig:Unc_numSims}.

For several of the parameters, $\Omega_m$, $\sigma_8$, $n_s$, $M_A$, $M_B$ and $\mathrm{ln} \sigma$, their posterior 1$\sigma$ standard deviations decrease with increasing number of training samples. 
For example, with the $\Omega_m$ parameter, when using only 100 sets of training simulations, the posterior standard deviation can be larger than 0.075. 
In contrast, the prior range of $\Omega_m$ is from 0.1 to 0.5. 
However, after increasing the number of training simulations to 3000, the posterior standard deviation of $\Omega_m$ drops to less than 0.025, more than a factor of 3. Afterwards, the standard deviation of $\Omega_m$ only reduces very mildly. The trends with the $\sigma_8$ and $M_A$ parameters are similar.

Very interestingly, with the $n_s$, $M_B$ and the $\mathrm{ln} \sigma$ parameters, their posterior 1$\sigma$ standard deviation appears to be stable when the training set size is smaller than $\sim 3000$, and only appear to improve after the size of the training set is larger than 3000, by which time, the posterior 1$\sigma$ standard deviations of the $\Omega_m$, $\sigma_8$ and $M_A$ parameters have reduced significantly. It appears that the MDN algorithm in our application first learns about the distribution of the $\Omega_m$, $\sigma_8$ and $M_A$ parameters, and only after their probability functions have been reasonably constrained, the algorithm starts to pick up additional information about the probability distributions of $n_S$, $M_B$ and $\mathrm{ln} \sigma$. In the future, it would be interesting to test whether or not such trends persist when changing the configuration of the MDN algorithm or when using other types of density estimators. 

Finally, for the $\Omega_b$ and $h$ parameters, their posterior standard deviations do not appear to change significantly with an increasing amount of training simulations. As we will discuss in the results sections, this is likely due to a lack of constraining power from the observables. Because the posterior standard deviations of all these five parameters, $\Omega_m$, $\sigma_8$ and $M_A$, $n_S$, $M_B$ and $\mathrm{ln} \sigma$, only appear to reduce very mildly after 10,000 simulations, we conclude that the default training sample used in this paper, which contains 150,000 sets of analytically-derived simulations, is sufficient for deriving representative posterior samples for the model parameters. 

The bias quantities for all of the parameter posterior samples appear to be stable for training set sizes 100 to 100,000. 
Therefore, we conclude that the training set size does not appear to be relevant for the parameters' biases, but is necessary to consider for accurately inferring the posterior distribution widths.
\subsection{Setup of the Markov Chain Monte Carlo Application}
\label{sec:mcmc_method}

We apply the Markov Chain Monte Carlo (MCMC) method to sample the posterior distribution of model parameters for the observable data vector. The process of this application is broadly illustrated in Figure~\ref{fig:sbi_block}. 

The application of MCMC often depends on using an explicit, analytical likelihood function. According to Bayes' theorem, a posterior probability function of a parameter set, $\theta$, given a set of observables $x$ can be derived from the likelihood function of the observables, $P(x|\theta)$, and the prior distribution of $\theta$, $P(\theta)$, as in 
\begin{equation}
    P(\theta|x) \propto P(x|\theta) P(\theta). 
\label{eq:bayes}
\end{equation}
MCMC can return a posterior sample of the model parameters by first proposing parameter values from the prior distribution $P(\theta)$ and updating the acceptance and rejections of these parameters according to the corresponding likelihood values of the observables $P(x|\theta)$. 
Typically, analyses of large-scale structure and galaxy clusters assume a Gaussian distribution for the likelihood function.
Under this approximation, the likelihood can be written as
\begin{equation}
    P(x| \theta) = \mathcal{N}(\bar{x}(\theta)-x, S(\theta)), 
\label{eq:n_like}
\end{equation}
where $\bar{x}(\theta)$ is the prediction for the observable data vector for  a given set of parameters $\theta$, and $S(\theta)$ is the Gaussian covariance matrix that describes the expected variance of the actual observable. \cite{2023MNRAS.520.6223P} have shown that this Gaussian likelihood function assumption  for galaxy cluster cosmology analyses is accurate even in the LSST era. Thus, we also implement a Gaussian likelihood assumption in this analysis.

To compute the observable prediction values for each model parameter set, we use the analytical forms of the halo mass function and the richness-mass relation, which were described in the previous sections. Specifically, the occurrence probability of a cluster with mass $M$, redshift $z$ and richness value $\lambda$ can be derived from the halo mass function $h(M, z)$ and the richness-mass relation as $\propto h(M, z) P(\lambda|M, z)$. Considering the halo densities and the volume of the simulation, we can derive the prediction for the total number of clusters and their average masses in a richness range $\Delta \lambda$ at redshift $z$ as 

\begin{equation}
\begin{split}
\overline N (\theta)_{\Delta \lambda, z} & = V_\mathrm{sim}  \int_{10^{14}}^{\infty} \mathrm{d}M \int_{\Delta \lambda} \mathrm{d} \lambda  ~ h(M, z | \theta)  P(\lambda|M, z, \theta),  \\
\overline{NM}(\theta)_{\Delta \lambda, z} &= V_\mathrm{sim}  \int_{10^{14}}^{\infty} \mathrm{d}M \int_{\Delta \lambda} \mathrm{d} \lambda   M
h(M, z | \theta) P(\lambda|M, z, \theta), \\
\overline{M}(\theta)_{\Delta \lambda, z} & = \overline {NM}(\theta)_{\Delta \lambda, z} / \overline{N}(\theta)_{\Delta \lambda, z}. 
\end{split}
\label{eq:cluster_counts}
\end{equation}
In the above equation, $V_\mathrm{sim}$ is the volume of the simulations, which is $1\mathrm{Gpc}^3h^{-3}$, and $h(M, z | \theta)$ is calculating the number densities of halos with mass value $M$ and redshift value $z$, for a cosmology parameter set $\theta$. Because the analytical simulations apply a halo mass threshold of $10^{14} M_\mathrm{\odot}/h$, the models above also incorporate a lower bound of $10^{14} M_\mathrm{\odot}/h$ for the halo mass distribution. Note that in the above set of equations,  the average mass of clusters in a richness range $\overline{M}(\theta)_{\Delta \lambda, z}$, is derived from the total mass of clusters $\overline{NM}(\theta)_{\Delta \lambda, z}$ in the range divided by the cluster number counts $\overline N (\theta)_{\Delta \lambda, z}$. 

To derive the covariance matrix  $S(\theta)$, we use the analytical simulation procedures described in the previous section. For each parameter set $\theta$, we derive 100 sets of analytical simulations, and compute the covariance of the cluster observables. 
We assume the covariance matrix to be diagonal (as the off-diagonal terms of the covariance is close to 0), and only consider the variance of each observable element among the 100 sets of simulations. For each MCMC run, we compute the covariance matrix once with the truth parameters of the corresponding test case. 

Finally, our MCMC procedure is implemented with the \text{EMCEE} \ sampler \citep{2013PASP..125..306F} and the \text{CosmoSIS} software package \citep{2015A&C....12...45Z}. For each posterior sample acquired from the MCMC method, we test the chain convergence with the Gelman-Rubin \citep{1992StaSc...7..457G} method implemented in the \text{ChainConsumer} package \citep{Hinton2016}. We also perform additional tests such as chain trace plotting. The MCMC posterior samples presented in this paper have been pruned of the first half of the run to exclude the ``burn-in" period. 

\begin{table*}
\begin{tabular}{| l | l |l|l|l|l|l|l|l|l|}
\hline
                               &                        & $\Omega_m$ & $\Omega_b$ & $h$ & $n_s$ & $\sigma_8$ & $M_\mathrm{A}$ & $M_\mathrm{B}$ & $\mathrm{ln} \sigma$ \\ \hline
                            &     Prior   & $[0.1, 0.5]$ & $[0.03, 0.07]$ & $[0.5, 0.9]$ & $[0.8, 1.2]$ & $[0.6, 1.0]$ & $[3.2, 5.2]$ & $[0.99, 1.49]$ & $[0.456, 0.756]$ \\ \hline
\multirow{4}{*}{1} & Truth          & 0.3385 & 0.0590 & 0.5168 & 0.8203 & 0.6401 & 4.5470 & 1.1067 & 0.5008                   \\ \cline{2-10} 
                               & Ana-SBI & 0.328$\pm$0.005 & 0.053$\pm$0.009 & 0.61$\pm$0.07 & 0.89$\pm$0.06 & 0.6339$\pm$0.0037 & 4.61$\pm$0.05 & 1.17$\pm$0.05 & 0.493$\pm$0.024                     \\ \cline{2-10} 
                               & MCMC    & 0.332$\pm$0.006 & 0.055$\pm$0.011 & 0.5855$\pm$0.0694 & 0.86$\pm$0.05 & 0.636$\pm$0.004 & 4.585$\pm$0.048 & 1.153$\pm$0.047 & 0.499$\pm$0.022                    \\ \hline
\multirow{4}{*}{2} & Truth             & 0.4292 & 0.0591 & 0.5028 & 0.8203 & 0.6327 & 3.6359 & 1.3316 & 0.5293                      \\ \cline{2-10} 
                               & Ana-SBI & 0.413$\pm$0.019 & 0.054$\pm$0.009 & 0.59$\pm$0.05 & 0.860$\pm$0.037 & 0.635$\pm$0.007 & 3.64$\pm$0.11 & 1.372$\pm$0.046 & 0.57$\pm$0.06                      \\ \cline{2-10} 
                               & MCMC    & 0.434$\pm$0.020 & 0.053$\pm$0.011 & 0.58$\pm$0.07 & 0.87$\pm$0.05 & 0.633$\pm$0.007 & 3.60$\pm$0.13 & 1.411$\pm$0.046 & 0.57$\pm$0.06                      \\ \hline
\multirow{4}{*}{ 3} & Truth           & 0.1652 & 0.0398 & 0.5001 & 0.8333 & 0.7410 & 3.8703 & 1.3639 & 0.5365                      \\ \cline{2-10} 
                               & Ana-SBI  & 0.156$\pm$0.007 & 0.053$\pm$0.008 & 0.66$\pm$0.09 & 0.95$\pm$0.08 & 0.759$\pm$0.013 & 3.81$\pm$0.10 & 1.32$\pm$0.06 & 0.58$\pm$0.06                      \\ \cline{2-10} 
                               & MCMC    & 0.156$\pm$0.007 & 0.055$\pm$0.010 & 0.66$\pm$0.11 & 0.94$\pm$0.09 & 0.755$\pm$0.012 & 3.84$\pm$0.10 & 1.34$\pm$0.06 & 0.54$\pm$0.06                      \\ \hline

\end{tabular}
\caption{\label{tbl:AnaTests} 
Inference of eight parameters on three test cases as shown in Figure~\ref{fig:contour_ana_ana}.
The test sets have different truth values of each parameter. 
For each test, we list their truth values,  and posterior mean and uncertainties from SBI trained on analytical simulations, and from the MCMC method. The prior ranges for the parameters are also listed, which is the same for all test cases and for all methods (SBI and MCMC) implemented in this paper.
}
\end{table*}
\section{SBI Application to Analytical Test Cases}
\label{sec:results}

\begin{figure*}
\begin{center}
\includegraphics[width=2.05\columnwidth]{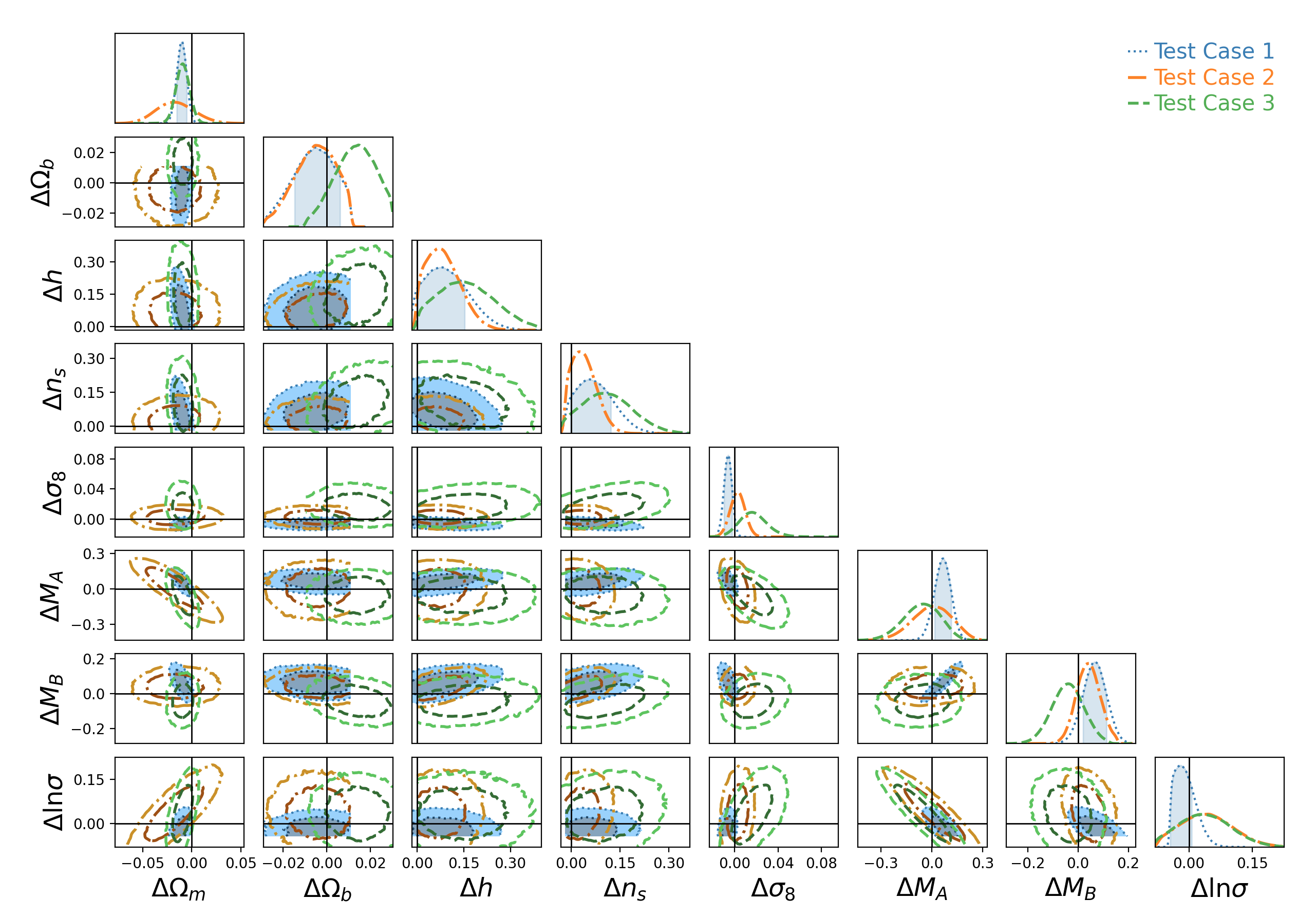}
    \caption{Corner plot showing the deviation of SBI posterior distributions for three test cases  listed in Table~\ref{tbl:AnaTests}. We show only the deviations between the parameter's posterior sample values and their truth values, given the large variations in the tests' truth values. The SBI method is trained and tested on the analytical simulations, and the contours represent the 1-sigma and 2-sigma confidence intervals for each test. The thin vertical and horizontal black lines indicate the truth values (zero deviations). They fall within the 2-sigma contour lines for all eight parameters, qualitatively suggesting that SBI method can return parameter truth values  within this uncertainty level.}
    \label{fig:contour_ana_ana}
\end{center}
\end{figure*}

\begin{figure*}
\begin{center}
\includegraphics[width=2.0\columnwidth]{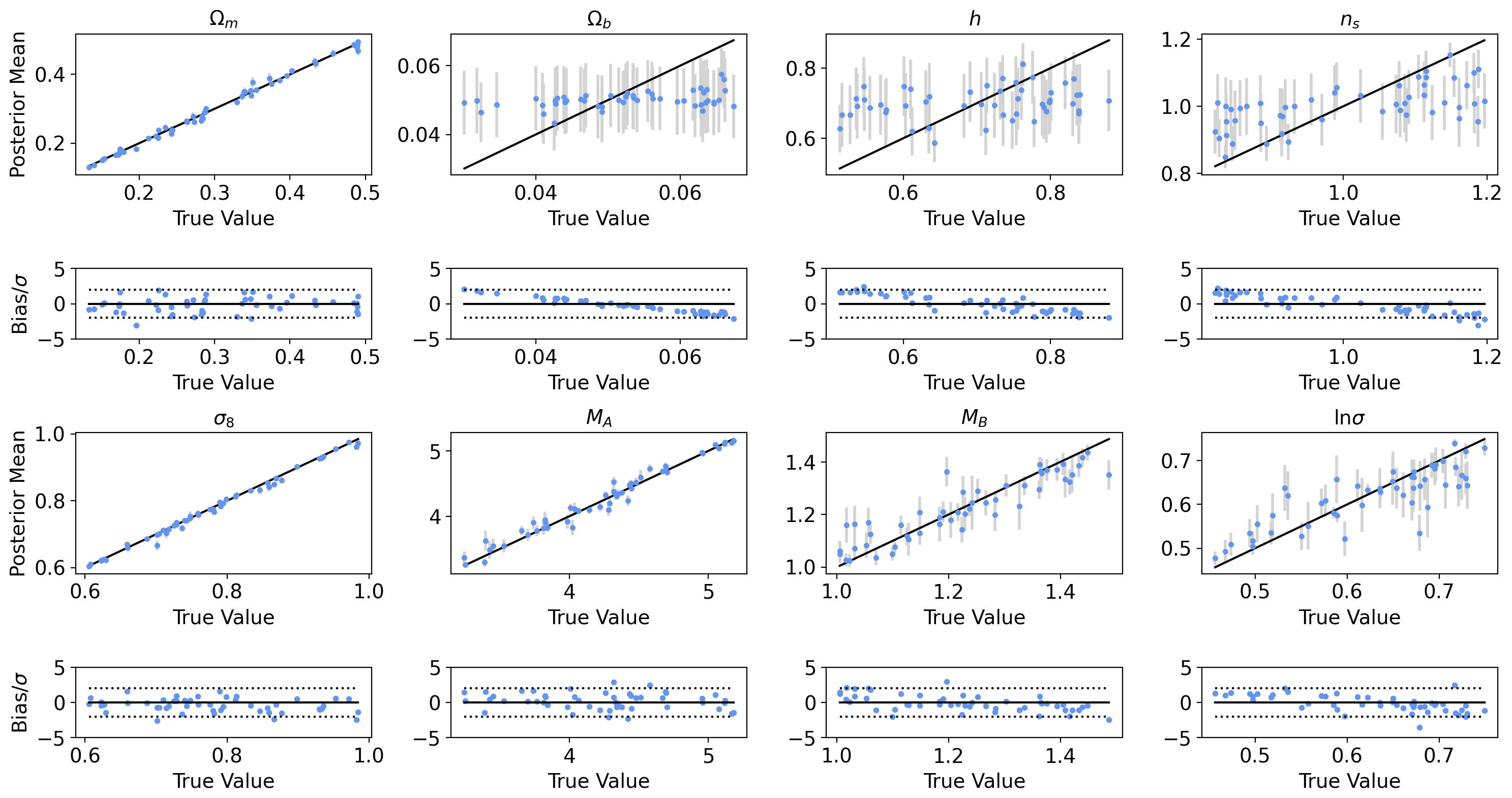}
    \caption{ A comparison between the truth values of each parameter and its posterior values derived from the SBI method,  for 50 sets of randomly selected test cases. 
    The bigger upper panels show the direct comparisons between the posterior means and uncertainties (standard deviation of the posterior sample) for each parameter. The blue circles represent the posterior means, the gray error bars represent the 1-$\sigma$ uncertainties (1 $\sigma$ standard deviation of the posterior sample) and the black lines represent the ideal case, in which the predicted and the truth values are equal. The smaller lower panels show the fractional biases between the posterior mean and the truth values, normalized by the posterior uncertainties. The black lines in these smaller panels indicate a 0-bias scenario, while the dotted lines indicate 2-$\sigma$ deviations.
    For both tests, for $\Omega_m$, $\sigma_8$, $M_A$, $M_B$ and $\mathrm{ln}\sigma$, the truth values are mostly recovered by the error bars and the biases are randomly distributed around zero. 
    For $n_s$, some correlation is observed between the truth and the predicted values. 
    For $\Omega_b$ and $h$, the predicted values are roughly constant, irrespective of the truth values, which leads to positive bias for smaller truth values of these parameters and vice versa. In addition, the Pearson Correlation $p$-value test shows no significant correlation between truth and posterior values for $\Omega_b$ and $h$.
    The trends indicate that the cluster observables we use for this study are likely not sensitive to  $\Omega_b$ and $h$ in the model. See Section~\ref{sec:1to1_bias_unc} for detailed discussions.}
    \label{fig:posterior_1to1}
\end{center}
\end{figure*}

\begin{figure}
\includegraphics[width=\columnwidth]{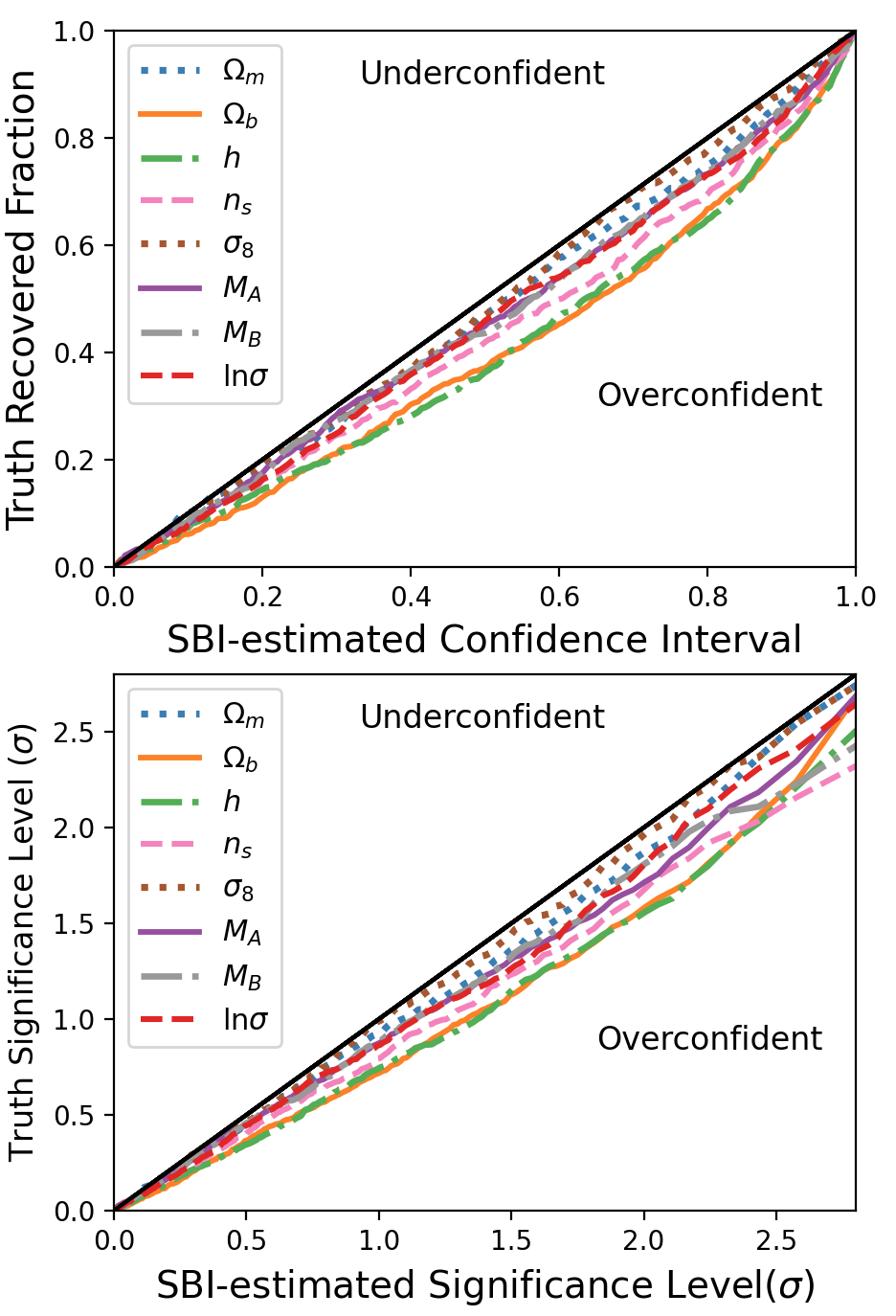}
    \caption{Coverage plot of the SBI posterior samples, with which we examine the statistical fidelity of the parameter confidence interval yielded by the SBI method. Upper Panel: This figure shows the distribution of the fraction of parameter truth values recovered on the $y$-axis vs the confidence interval on the $x$-axis estimated by SBI. The solid black line represents the ideal scenario, where the fraction of truth values recovered coincides with the SBI-estimated confidence interval. The colored lines with different line styles are the recovery rate for each of the 8 parameters. The SBI posterior tends to be overconfident, and especially so for the $\Omega_b$, $h$, and $n_s$ parameters. Lower Panel: The equivalent significance level for the coverage plot in the upper panel. The fraction of parameter truth values are converted into a significance level estimation assuming a Gaussian posterior distribution  --- e.g., 2 $\sigma$ significance level indicates 95\% fraction of truth values recovered. This recovered significance level is shown on the $y$-axis, with the equivalent significance level for the SBI confidence interval (assuming a Gaussian distribution as well) shown on the $x$-axis. Similar with the figure above, the SBI method is generally overconfident, which indicates that SBI tends to under-estimate parameter error bars. See Section ~\ref{sec:ana_coverage} for more detailed discussions.
     }
    \label{fig:coverage_anatest}
\end{figure}

In this section, we evaluate the posterior results acquired with the aforementioned SBI method for a set of test observables from the analytical simulations. In this and next Section, we focus on SBI method trained with the analytically generated simulations, with test cases also from the analytical simulations. We aim to test both the accuracy and precision of the posterior results with these carefully controlled simulations. 
The rest of this section details the findings with the evaluation methods. 

\subsection{Distribution of Parameter Posterior Samples}

In this section, we present the parameter posterior samples for three test cases from the SBI method and discuss their implications. 

Figure \ref{fig:contour_ana_ana} presents a ``corner'' plot of the  posterior parameter samples. The posterior sample for each test case is the joint distribution of eight parameters, and the truth values for  each of the eight parameters vary between the three test cases. To enable the presentation of the three test cases in the same corner plot, we present the deviations of the posterior parameter values from their truth values, and these truth values are listed in Table~\ref{tbl:AnaTests}. For the three test cases, the deviations of the eight parameters are consistent with 0 within the $2\sigma$(standard deviation) ranges. This result qualitatively demonstrates that truth values of the parameters can be reasonably expected within the confidence interval of the SBI posterior results.  We also see strong correlations between some of the parameter pairs, $h$ and $n_s$, $\Omega_m$ and $\sigma_8$, $M_A$ and $M_B$, $M_A$ and $\mathrm{ln}\sigma$. These kinds of correlation/degeneracy, especially between $\Omega_m$ and $\sigma_8$, $M_A$ and $M_B$, $M_A$ and $\mathrm{ln}\sigma$ have been observed in other galaxy cluster cosmology analyses \citep{2020PhRvD.102b3509A}, and additional constraints from other types of cosmological observations, such as the cosmic microwave background \citep{2020A&A...641A...6P}, Type Ia supernovae \citep{2024arXiv240102929D}, and baryon acoustic oscillations analyses \citep{2024arXiv240403002D}, will be needed to break the degeneracy.

The posterior confidence intervals for the parameters vary between each test case. For example, the uncertainty of $\Omega_m$ on tests 1 and 3 is smaller than the uncertainty acquired on test 2. In contrast, for $\sigma_8$, the posterior uncertainties are smallest on test 1, higher on test 2 and the highest on test 3. The result is not surprising given their different truth parameter values and observable data vectors -- we expect the parameter posterior distribution $P(\theta|x)$ to vary depending on the value of $x$, for a fixed $P(\theta, x)$ joint distribution. We will further compare the derived posterior uncertainties to those derived from the MCMC method in Section ~\ref{sec:mcmc}.

\subsection{Bias and Uncertainties} 
\label{sec:1to1_bias_unc}

We examine the posterior distributions from the SBI method against the truth on a test set of 50 analytical simulations. 
We study the bias, which is the difference between a parameter's  posterior sample values and the parameter's truth value.

The upper panels of Figure~\ref{fig:posterior_1to1} show the mean parameter values of the posterior samples (accompanied by the posterior uncertainties) versus their truths. For the $\Omega_m$, $\sigma_8$, $M_A$, $M_B$, and $\mathrm{ln}\sigma$ parameters, their mean posterior values tightly correlate with their truth values on a 1-1 relation. However, for $\Omega_b$, the mean of the posterior values appear to be constant across the truth value range. 
The galaxy cluster number counts and average masses are not known to be sensitive to or can effectively constrain $\Omega_b$ in the $\mathrm{\Lambda CDM}$ cosmology model because baryon and dark matter behave similarly on the physical scale of a galaxy cluster. 
Similarly, the posterior mean values of $h$ also appears to be constant, regardless of the truth value: the cluster observables are constructed to be independent of the universe's geometry. 
Interestingly, for $n_s$, the posterior mean values slightly correlate with the truth values, but the correlation is much weaker compared to that for  $\Omega_m$ and $\sigma_8$.
This parameter mainly affects our cluster observables through  the matter power spectrum in the $\mathrm{\Lambda CDM}$ model, but its effect is less dramatic than $\Omega_m$, $\sigma_8$, thus explaining the much weaker constraining power we have on this parameter. The Pearson correlation $p-$value test confirms our finding of no significant correlation between the truth and posterior values for $\Omega_b$ and $h$. The Pearson correlation test confirms a correlation between the $n_s$ posterior values and its truth values, but the $p$ value is much smaller than that for the $\Omega_m$, $\sigma_8$, $M_A$, $M_B$ and $\mathrm{ln}\sigma$, indicating weaker correlations. 

In observational studies, the posterior distributions are often quoted by their mean and uncertainty values, and tensions between different results are evaluated by the differences in mean values, normalized by their uncertainties, known as  significance level. We also examine the significance level of the deviations between the posterior mean and the truth, normalized by the posterior uncertainties, shown in the smaller lower panels in Figure~\ref{fig:posterior_1to1}. For $\Omega_m$, $\sigma_8$, $M_A$, $M_B$ and $\mathrm{ln}\sigma$, the inferred posterior mean values mostly scatter within 2.0$\sigma$ of the truth value, indicating that the truth values of the parameters are usually well represented by their 2.0$\sigma$ significance levels.  

\subsection{Posterior Coverage Analysis} 
\label{sec:ana_coverage}

In this section, we investigate the fidelity of the confidence interval of SBI posterior samples, making use of a ``coverage plot''. This test aims to answer the question ``What fraction of times the truth value is recovered by the SBI posterior samples within a specific confidence interval''?  During this test, we acquire posterior samples for 1000 test cases. For each test case, we decide a posterior confidence range corresponding to a probability percentage of $p$. Then, for the 1000 test cases, we record the fraction of them whose truth values fall into the $p$ posterior confidence range. These fractions versus the varying values of $p$ are shown in the upper panel of Figure~\ref{fig:coverage_anatest}. In an ideal scenario, the probability distribution of the truth parameter values are perfectly recovered by the posterior sample distribution, and the fractions of truth values falling into a probability $p$ confidence range should also be $p$. If the fractions tend to be lower than the values of $p$, then the method is considered to be overconfident; if the fractions tend to be higher than the values of $p$, then the method is considered underconfident.  Ideally, SBI should be neither overconfident nor underconfident, but being overconfident is usually the lesser desired situation. 

The coverage plot (Figures \ref{fig:coverage_anatest}) reveals that the SBI posterior estimation is close to being faithful, but unfortunately does have a tendency of being overconfident. For five parameters, $\Omega_m$, $\sigma_8$, $M_A$, $M_B$ and $\mathrm{ln}\sigma$, the relation between the truth fractions and the confidence levels are close to 1:1, with a maximum deviation of around 7.8\% with the $\mathrm{ln}\sigma$ parameter. This indicates that the posterior probability distribution from SBI tend to reflect the posterior uncertainties of these parameters. For $\Omega_b$, $h$ and $n_s$, the method is overconfident. However, as discussed in the previous section, the SBI posterior sampler returns a Gaussian-like distribution for these parameters even though their truth value distributions are flat, and thus tend to under-estimate the frequencies of these parameters appearing towards the edges of their prior ranges. 

We further examine whether or not the 1$\sigma$ standard deviation often quoted as ``error bars'' (or uncertainties) are accurately recovered by the SBI. We aim to test the question ``Are the SBI-estimated error bars too small or too large?'' We convert the frequencies and confidence probabilities in the upper panel of Figure~\ref{fig:coverage_anatest} into significance levels quantified as error bars, assuming a Gaussian distribution of the posterior probability function. Under the Gaussian assumption, the 68\% frequency and confidence interval would correspond to a 1$\sigma$ error bar, while 95\% frequency and confidence interval would correspond to a 2$\sigma$ error bar. A too-small or too-large error bar would carry consequences for how we interpret parameter estimation differences or tensions between different analyses. For example, if an SBI-estimated 1$\sigma$ error bar is only 50\% of the truth error bar, a 2$\sigma$ parameter tension between the SBI-based result and another analysis will only amount to a 1$\sigma$ difference, which is often well tolerated in the cosmology community.  In addition, this error bar conversion of the coverage plot is more sensitive to the distribution of the frequency and confidence intervals towards the low-probability region, probing the credibility of the SBI method in terms of quantifying outlier distributions. 

With this conversion, shown in the lower panel of Figure~\ref{fig:coverage_anatest}, we again find the SBI method to perform overconfidently. In the best-performing scenario, the estimated 1$\sigma$ error bar from SBI ($x-$axis of the figure) corresponds to a truth error bar of 0.99$\sigma$ for $\sigma_8$, while the 2$\sigma$ significance interval corresponds to a truth error bar of 1.97$\sigma$. In a worse-performing scenario, the estimated 1$\sigma$ error bar from SBI ($x-$axis of the figure) corresponds to a truth error bar of 0.88$\sigma$ for $M_A$, while the 2$\sigma$ significance interval corresponds to a truth error bar of 1.72$\sigma$. 
These results again demonstrate the need for testing and calibration before interpreting posterior parameter tensions acquired from the SBI method.

\section{SBI Comparison with the MCMC Method}\label{sec:mcmc}

\begin{figure*}
\includegraphics[width=2.0\columnwidth]{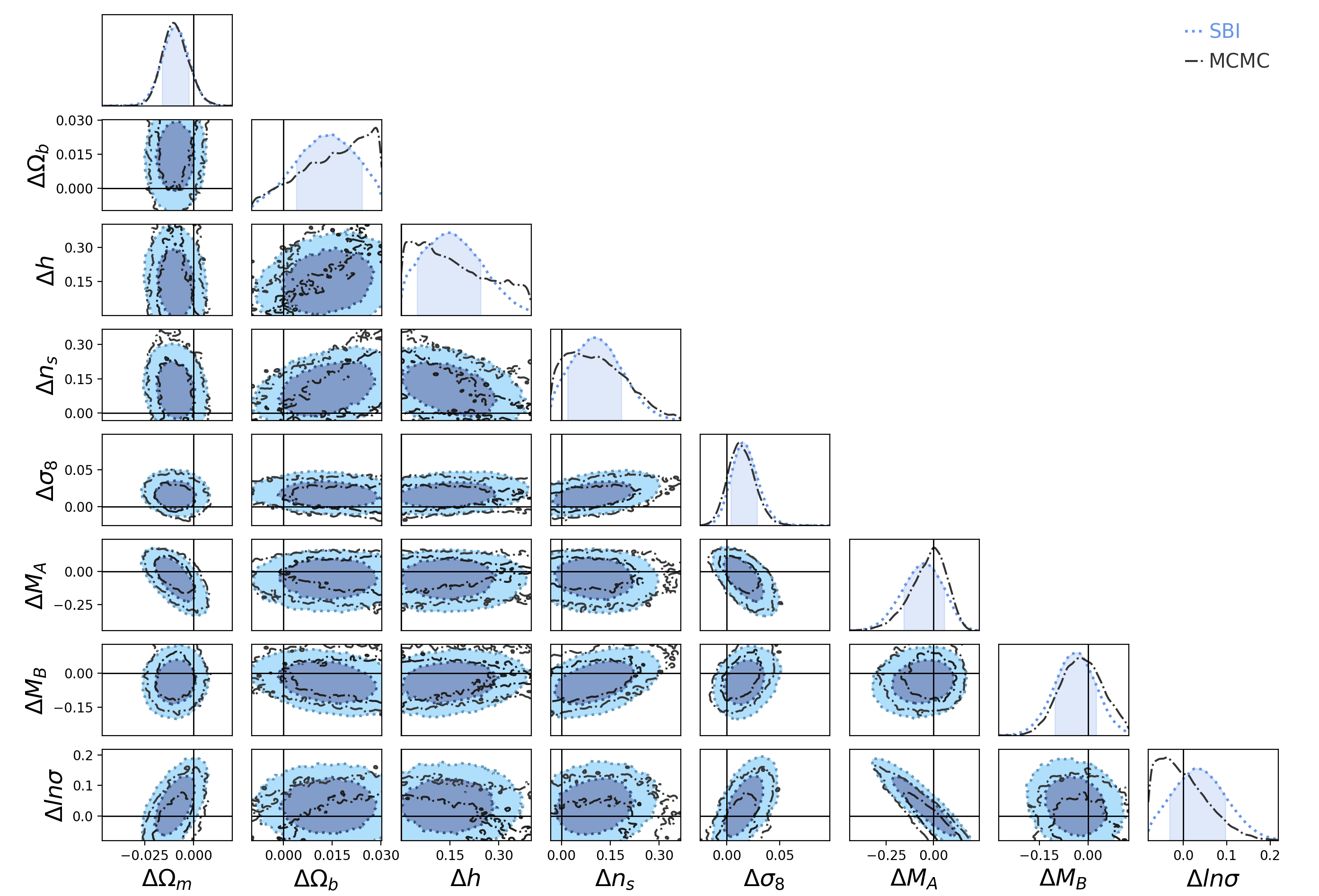}
    \caption{Corner plot showing the parameter posterior distributions for a single test case (Test Case 3 in Table~\ref{tbl:AnaTests}) with the SBI method and the MCMC method. For parameters $\Omega_m$,  $\sigma_8$, $M_A$, $M_B$, and $\mathrm{ln} \sigma$, the posterior distributions from SBI and MCMC are consistent. For the $\Omega_b$, $h$ and $n_s$ parameters which the cluster observable data vector do not strongly constrain, the posterior samples from SBI and MCMC have notably different shapes, which likely reflect the shortcomings of each method. See Section~\ref{sec:mcmc} for a more detailed discussion.}
    \label{fig:3methods_mcmc}
\end{figure*}

\begin{figure*}
\includegraphics[width=2.0\columnwidth]{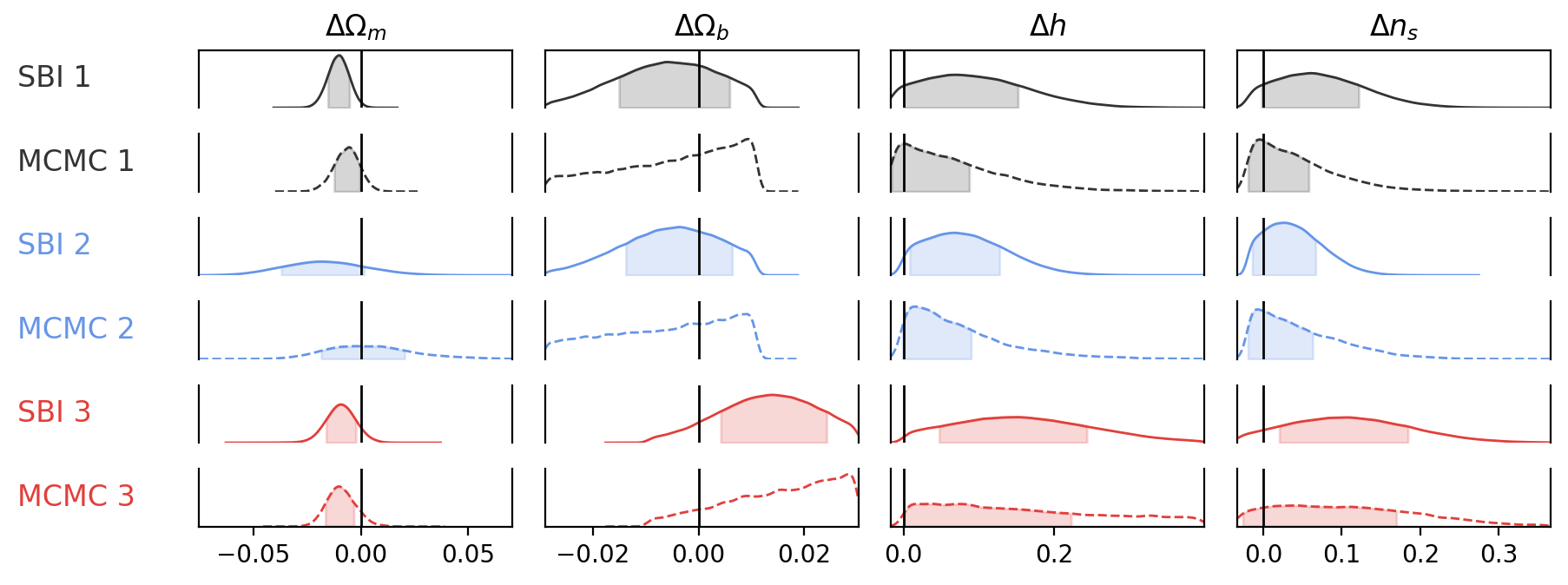}
\includegraphics[width=2.0\columnwidth]{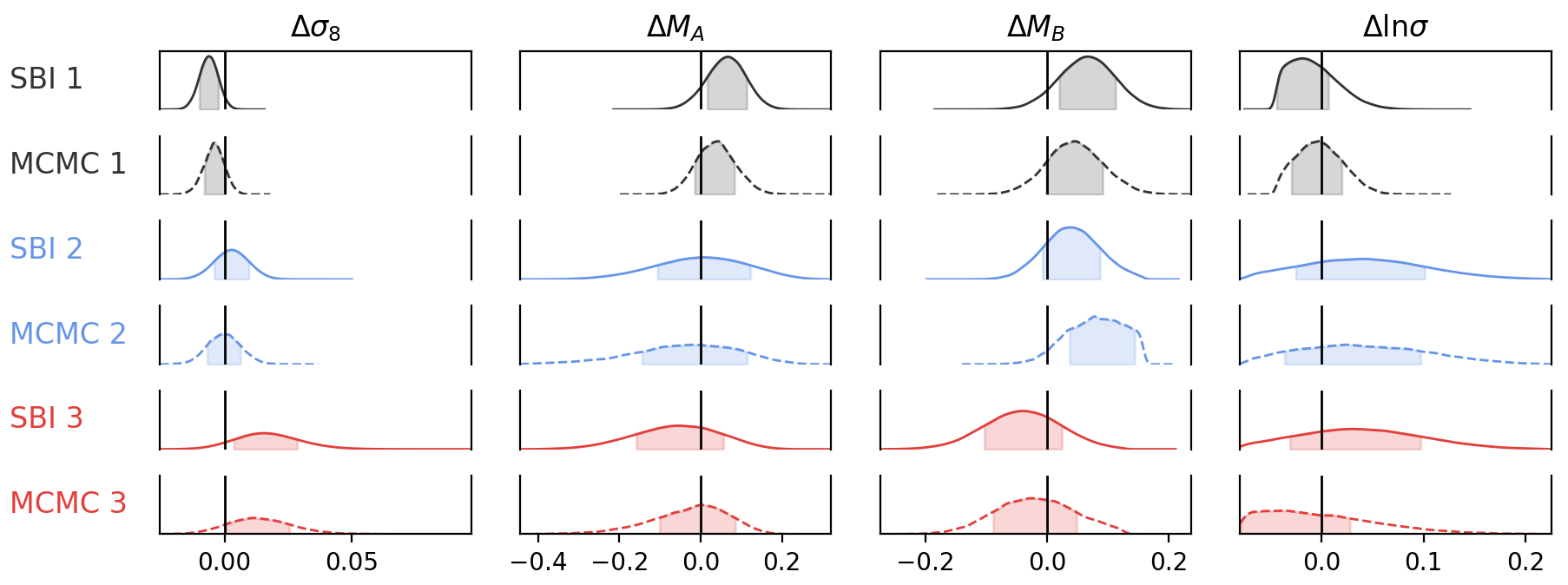}
    \caption{Column plot showing the marginalized posterior distributions from both the SBI and MCMC methods for three test cases. For each of the test cases, the MCMC and SBI posterior distributions are consistent within 1$\sigma$ (shaded region) for parameters $\Omega_m$,  $\sigma_8$, $M_A$, $M_B$, and $\mathrm{ln} \sigma$. For $\Omega_b$, $h$ and $n_s$, the posterior samples from SBI and MCMC have notably different shapes. See Section~\ref{sec:mcmc} for a more detailed discussion. Results are also listed in Table~\ref{tbl:AnaTests}.}
    \label{fig:3methods_columns}
\end{figure*}

\begin{figure*}
\includegraphics[width=1.8\columnwidth]{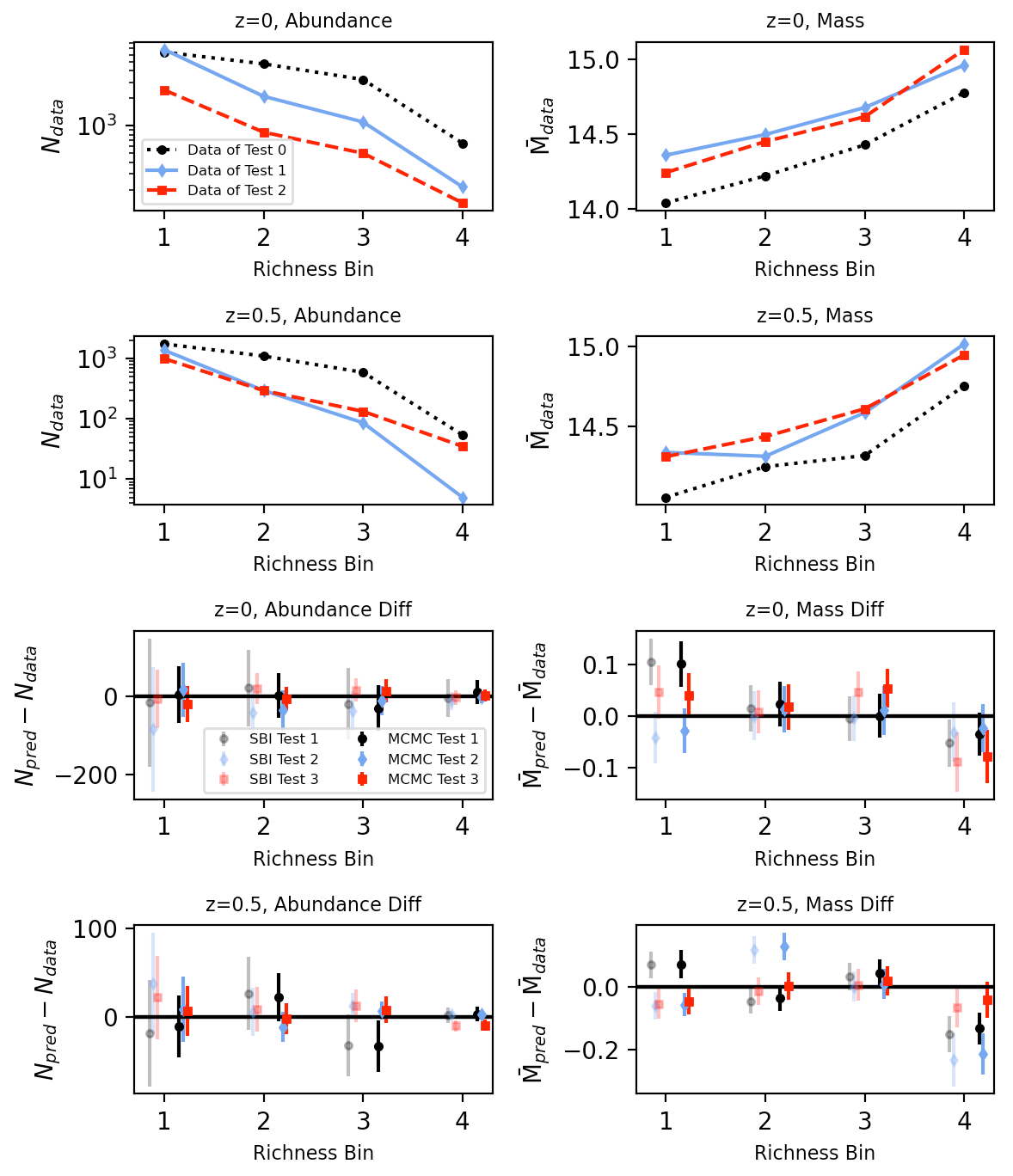}
    \caption{
    Cluster abundances (left column) and masses (right column) predicted using the posterior samples derived by the SBI and MCMC methods. 
    The upper two rows show the truth data vectors in the three test cases, while the lower two show the differences between the posterior predictions and the truth data vectors. The posterior predictions from the SBI and MCMC methods (the posterior prediction values are slightly offset along the $x-$axis direction to avoid overlapping data points) are highly consistent with each other. The posterior predictions on the cluster counts (lower left panels) are highly consistent with the truth data vector component, while the deviations between the posterior predictions and the average masses are more pronounced. See Section~\ref{sec:mcmc} for more detailed discussions. }
    \label{fig:MCMCSec_posterior_precitions}
\end{figure*}

In this section, we compare the SBI method to the MCMC method; the latter has been widely applied in cluster cosmology analyses. 
As previously mentioned, the MCMC implementation shares a set of physical model ingredients with the construction of the analytical simulations used in the SBI method, which includes sampling Poissonian statistics from a halo mass function, and probabilistically modeling the relation between cluster richness and mass. 
With these models, the posterior probability functions for the parameters are determined, but SBI and MCMC differ in the algorithms they use to approximate the posterior functions. SBI  uses a training set that contains information about the probability distribution of the observables, and an MDN-based machine learning method to approximate the parameter-observable probability function. The faithfulness of the SBI posterior samples in reflecting the parameter's real posterior probability function will depend on the accuracy of the MDN implementation. Previously, we have already observed that the SBI posterior samples are not faithful in reflecting the posterior probabilities of $\Omega_b$, $h$ and $n_s$, potentially revealing a failure in using MDN to approximate their probability functions. However, the algorithm appears to provide highly faithful posterior samples for the other five parameters: $\Omega_m$, $\sigma_8$, $M_A$, $M_B$ and $\mathrm{ln} \sigma$. 

On the other hand, our implementation of MCMC uses a Gaussian function to approximate the likelihood function of the observables. The faithfulness of the posterior samples in representing the parameter's real posterior probability function will depend on the accuracy of the Gaussian approximation of the likelihood function as well as the implementation of the MCMC process. 

With this reasoning, if the MCMC and SBI methods are both well calibrated, we expect MCMC and SBI to yield highly consistent posterior distributions for the five parameters: $\Omega_m$,  $\sigma_8$, $M_A$, $M_B$ and $\mathrm{ln}\sigma$. In this section, we perform MCMC analysis for some test cases and compare the posterior samples to those derived from the SBI method. We chose to analyze only three test cases due to the much longer time needed to derive the MCMC posterior samples. We compare the posterior samples from the MCMC and SBI methods on these three test cases, and also perform posterior predictive checks. 

\subsection{Comparison of the Parameter Posterior Samples}

Figure~\ref{fig:3methods_mcmc} shows a corner plot of the posterior sample for Test Case 3 in Table~\ref{tbl:AnaTests}. 
For parameters $\Omega_m$, $\sigma_8$, $M_A$ and $M_B$, the posterior sample distributions from SBI and MCMC are highly consistent. For $\mathrm{ln}\sigma$, the MCMC method yields posterior samples that are systematically shifted compared to the SBI result, but the mean and median of their posterior samples are still within 1$\sigma$ of each other. 
The MCMC posteriors for $\Omega_b$, $h$, and $n_s$ are less informative, and they have different shapes with the SBI result.
As previously noted, the SBI posterior distribution of these parameters tend to have a Gaussian shape, likely due to the SBI probability functions being modeled by multi-Gaussian distributions. On the other hand, the MCMC method yields posterior samples with a broad spread, skewed towards either end of the prior range. 
In addition, the MCMC Gelman-Rubin test shows that the MCMC posterior sampling for $\Omega_b$ has not converged, likely caused by a lack of constraining power from the observables for $\Omega_b$.

We further examine the marginalized posterior distributions of parameters from SBI and MCMC using ``Column''
plots (Figure~\ref{fig:3methods_columns}) for the three test cases in Table~\ref{tbl:AnaTests} (including the one already presented in Figure~\ref{fig:3methods_mcmc}). Similar trends are observed again for the SBI and MCMC method:  MCMC and SBI yield highly consistent results for $\Omega_m$,  $\sigma_8$, $M_A$, $M_B$. The exception is the $\mathrm{ln}\sigma$ posterior distributions, where the MCMC and SBI results are highly consistent for two test cases. It is possible that the slight shift between MCMC and the SBI constraints for $\mathrm{ln}\sigma$ in the Figure~\ref{fig:3methods_mcmc} test case is an outlier. For $\Omega_b$, $h$, and $n_s$ we again observe a similar trend --- that the SBI posterior distributions tend to appear Gaussian-like, while the MCMC distributions tend to have broad spreads, skewing towards one end of the value range.

\subsection{Comparison of Posterior Predictions} \label{sec:mcmc_PP}

We make posterior predictions of the observable data vector for the MCMC and SBI methods to explore their differences on the test cases. For each test case, we draw one set of parameter values from the posterior samples acquired with either the MCMC or SBI method, and then generate a set of observables -- the cluster number counts and average masses in four richness bins at redshift 0 and redshift 0.5. For the SBI method, this means re-running the observable simulation code with one set of model parameters drawn from the SBI posterior samples. For the MCMC method, this means recomputing the model observable prediction with one set of model parameters from the MCMC posterior samples using Equations~\ref{eq:cluster_counts}. We repeat the process for  100 sets of parameter values and make 100 sets of simulated observable data vectors, known as ``posterior predictions'', for the SBI and MCMC method respectively.

Figure~\ref{fig:MCMCSec_posterior_precitions} shows these posterior predictions for the cluster number counts and average masses in the richness bins at two redshifts for the three test cases in Table~\ref{tbl:AnaTests}. We show the mean values for the observable predictions based on the 100 sets of posterior parameter values, as well as their fluctuations. The SBI posterior predictions tend to have large dispersion than the predictions from the MCMC method. This is because the simulations from the SBI method fully simulate the noises in the observables including Poissonian fluctuations in the cluster counts observable and noises in the mass observables.  The MCMC predictions only simulate the mean values of the observables without measurement noises, but the measurement noises are captured by the covariance matrix used in the MCMC analysis. Thus, for the MCMC analysis, we indicate the fluctuation level of the MCMC posterior prediction with the square-root of the corresponding observable element in the covariance matrix. The SBI posterior prediction's dispersion (1$\sigma$ standard deviation) is highly consistent with this covariance matrix estimation used in the MCMC analysis. 

We examine the differences between the true observable data vectors and their posterior predictions.
In ideal situations, the posterior predictions should fluctuate around the truth observable data vector, which is indeed the case for the cluster number counts component (at either redshift 0 or redshift 0.5) for both the SBI and MCMC posterior samples (left column of Figure~\ref{fig:MCMCSec_posterior_precitions}).  
However, for the average mass observable component (Figure~\ref{fig:MCMCSec_posterior_precitions}; right column), deviations between the posterior predictions and the true observables are more noticeable --- sometimes more than $1\sigma$ deviations, but generally still consistent within $3\sigma$. 
One potential explanation is that the posterior distributions are more sensitive to fluctuations in the cluster counts observables than the mass observables. 
We test this hypothesis by perturbing either the cluster counts or the average masses in the test cases by adding a random noise (See the Appendix~\ref{app:sec:sensitivity} for more details), and find that perturbing the cluster counts  sometimes shift only the posterior distribution of $\Omega_m$ and $\sigma_8$ out of the five informative parameters, while perturbing the cluster masses tend to cause shifts in all of the five informative parameters $\Omega_m$,  $\sigma_8$, $M_A$, $M_B$, and $\mathrm{ln}\sigma_0$. The entanglement between the five parameters for predicting the masses potentially means that a higher degree of deviations in the mass observable component can be tolerated by the models.

The MCMC and SBI posterior predictions are highly consistent with each other for all three test cases. 
The difference in the posterior predictions between the methods is less than the bias that each method exhibits.
This high level of consistency means that any differences in their posterior parameter samples (discussed in the previous sub-section) have negligible effects on their predicted observables. 
Based on posterior predictive checks alone, neither method exhibits a relative advantage.

\section{Application to Quijote Simulation Observables}
\label{sec:result_quijote}

\begin{figure*}
\includegraphics[width=2.0\columnwidth]{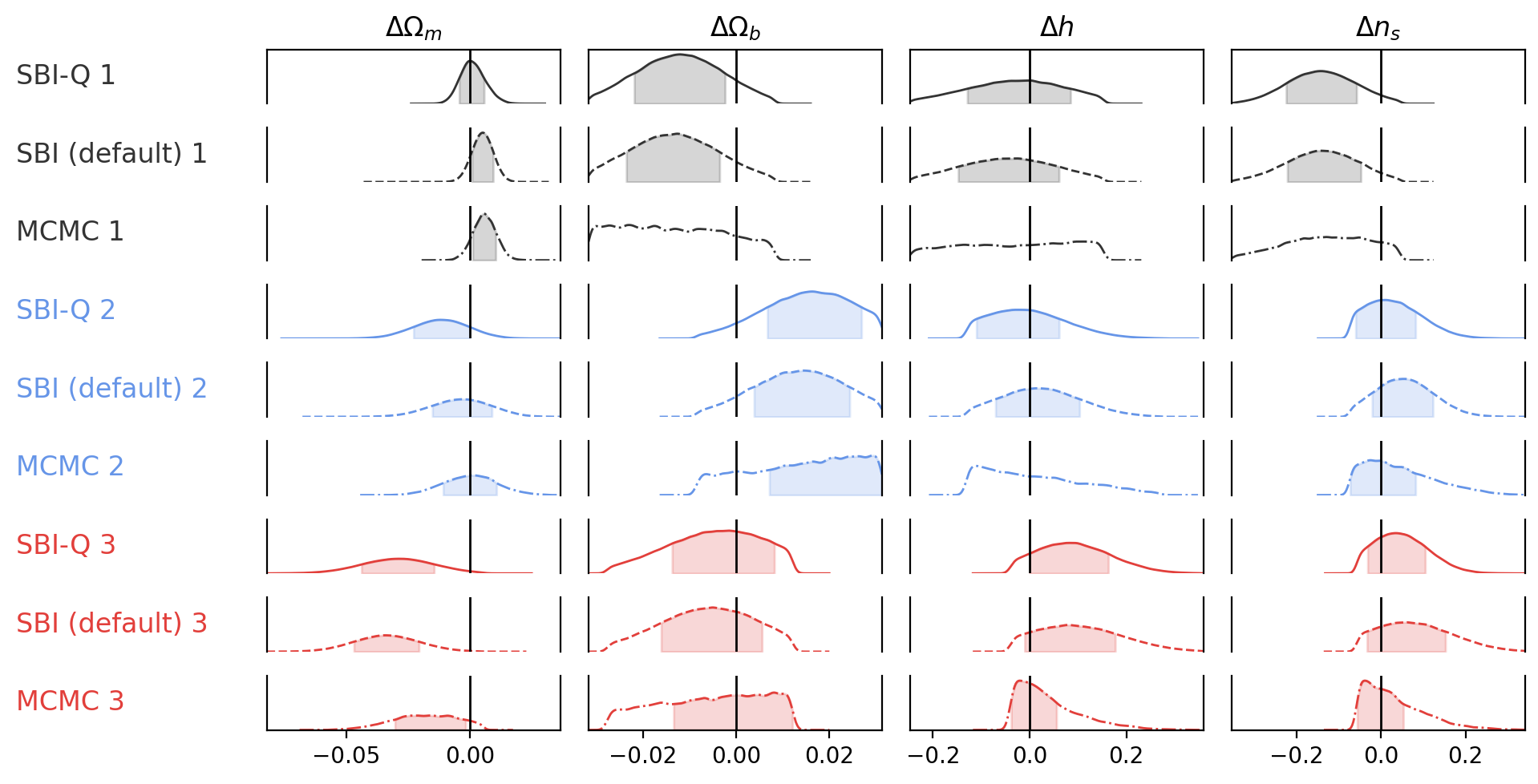}
\includegraphics[width=2.0\columnwidth]{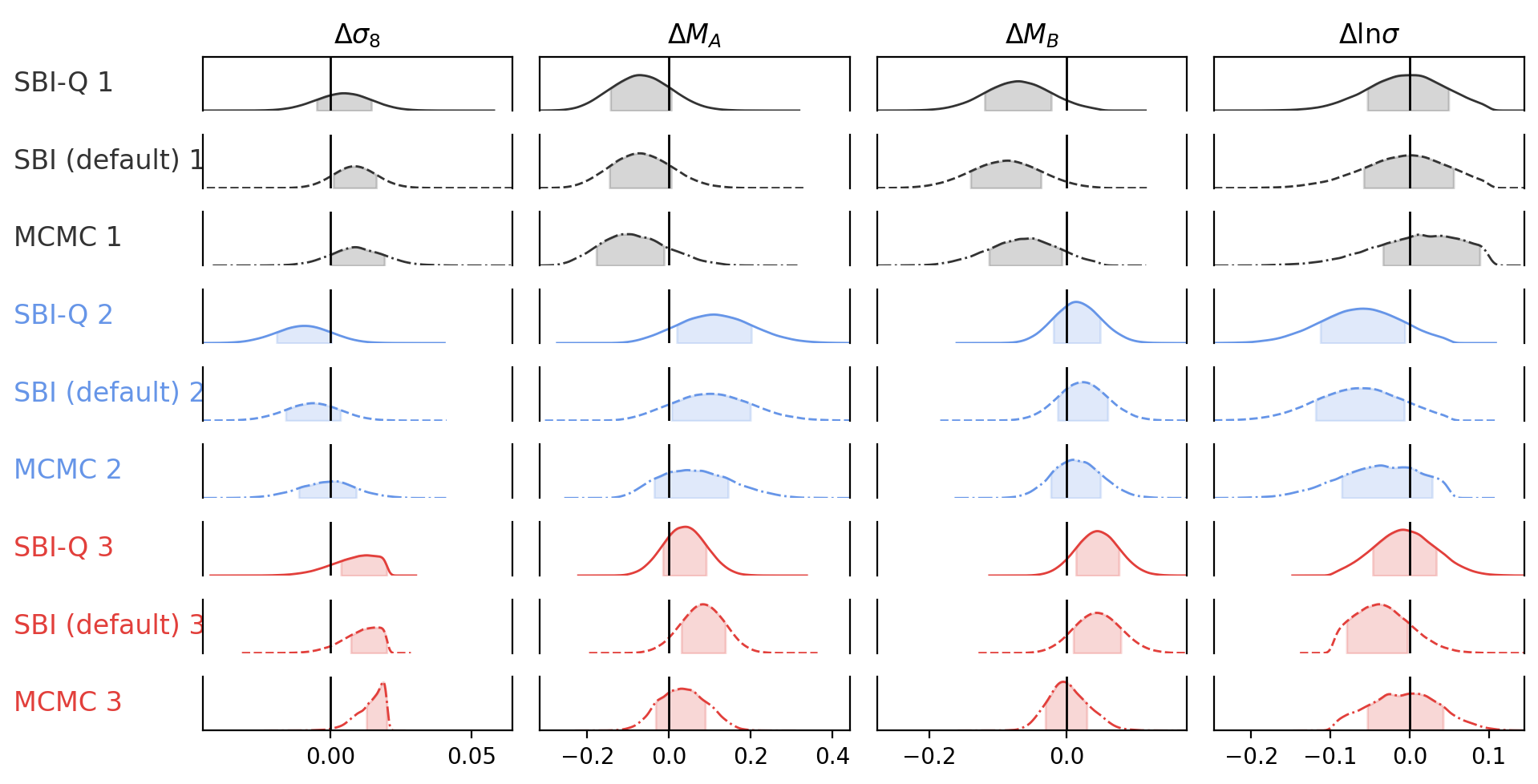}
    \caption{Column plot showing the marginalized posterior distributions for the three test cases from Quijote simulations, from the default SBI and MCMC methods, as well as from the SBI method trained on Quijote simulations (SBI-Q). The truth values of the parameters are recovered by all three methods within the 2-$\sigma$ level. 
    Results from the SBI methods trained on analytical simulations tend to align with those from the MCMC method for $\Omega_m$, $\sigma_8$, $M_A$, $M_B$, and $\mathrm{ln}\sigma$.
    For $\Omega_b$, $h$, and $n_s$, results from the SBI methods trained on analytical simulations and Quijote simulations tend to align.
    See Section~\ref{sec:qui_posterior} for a more detailed discussion.}
    \label{fig:Qui3methods_columns}
\end{figure*}

\begin{figure*}
\begin{center}
\includegraphics[width=2.1\columnwidth]{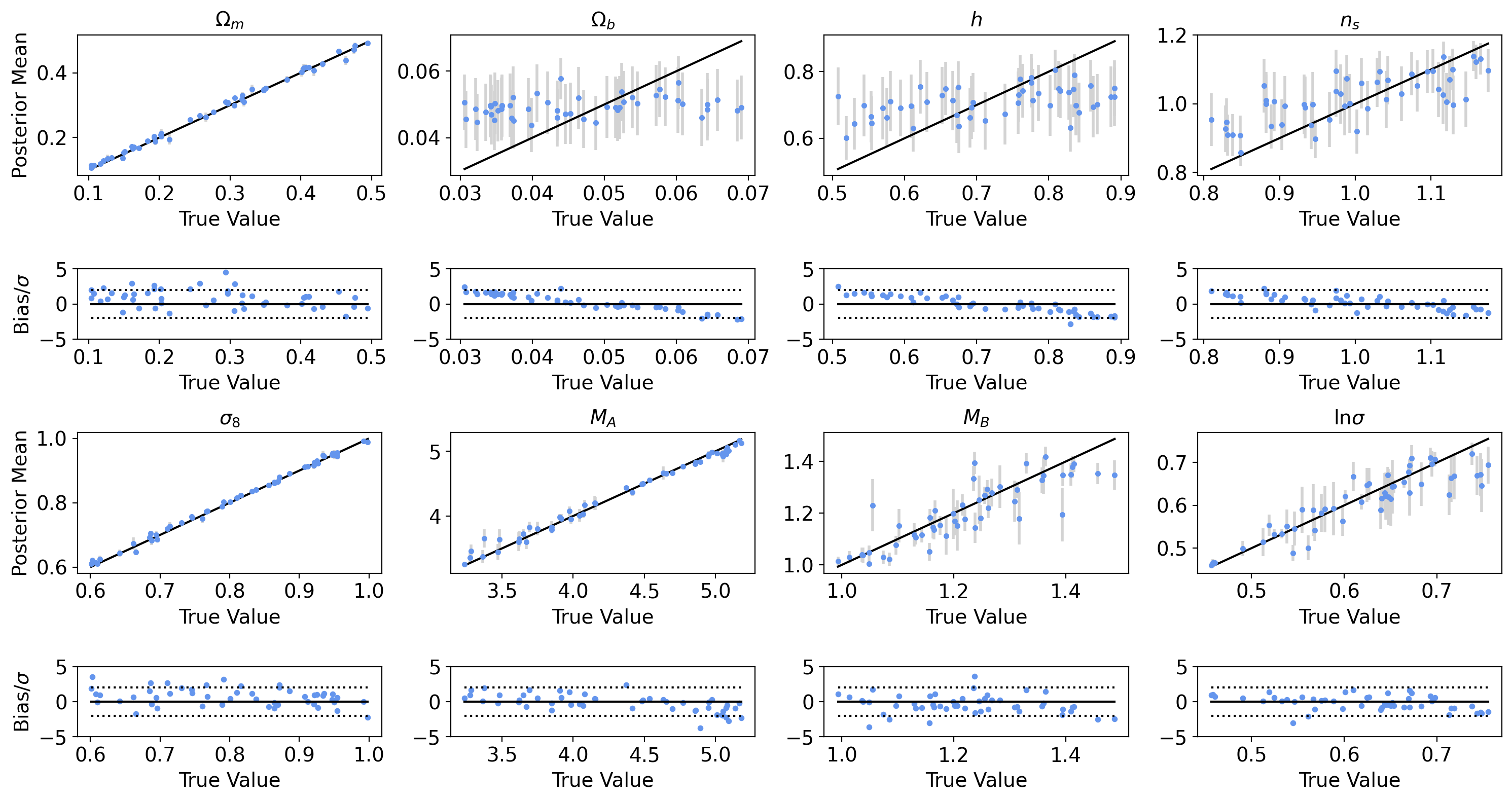}
    \caption{ 
    A comparison between the truth value of each parameter and its  posterior sample values derived from the default SBI method trained with analytical simulations, for 50 sets of randomly selected test cases from the Quijote simulations. 
    Among the five parameters that the observables have constraining power for ($\Omega_m$, $\sigma_8$, $M_A$, $M_B$, and $\mathrm{ln}\sigma$), the SBI method tends to yield biased estimations for $\Omega_m$ and $\sigma_8$. 
    See Section~\ref{sec:qui_test} for a more detailed discussion.}
    \label{fig:posterior_1to1_Qui}
\end{center}
\end{figure*}

\begin{figure*}
\includegraphics[width=2.0\columnwidth]{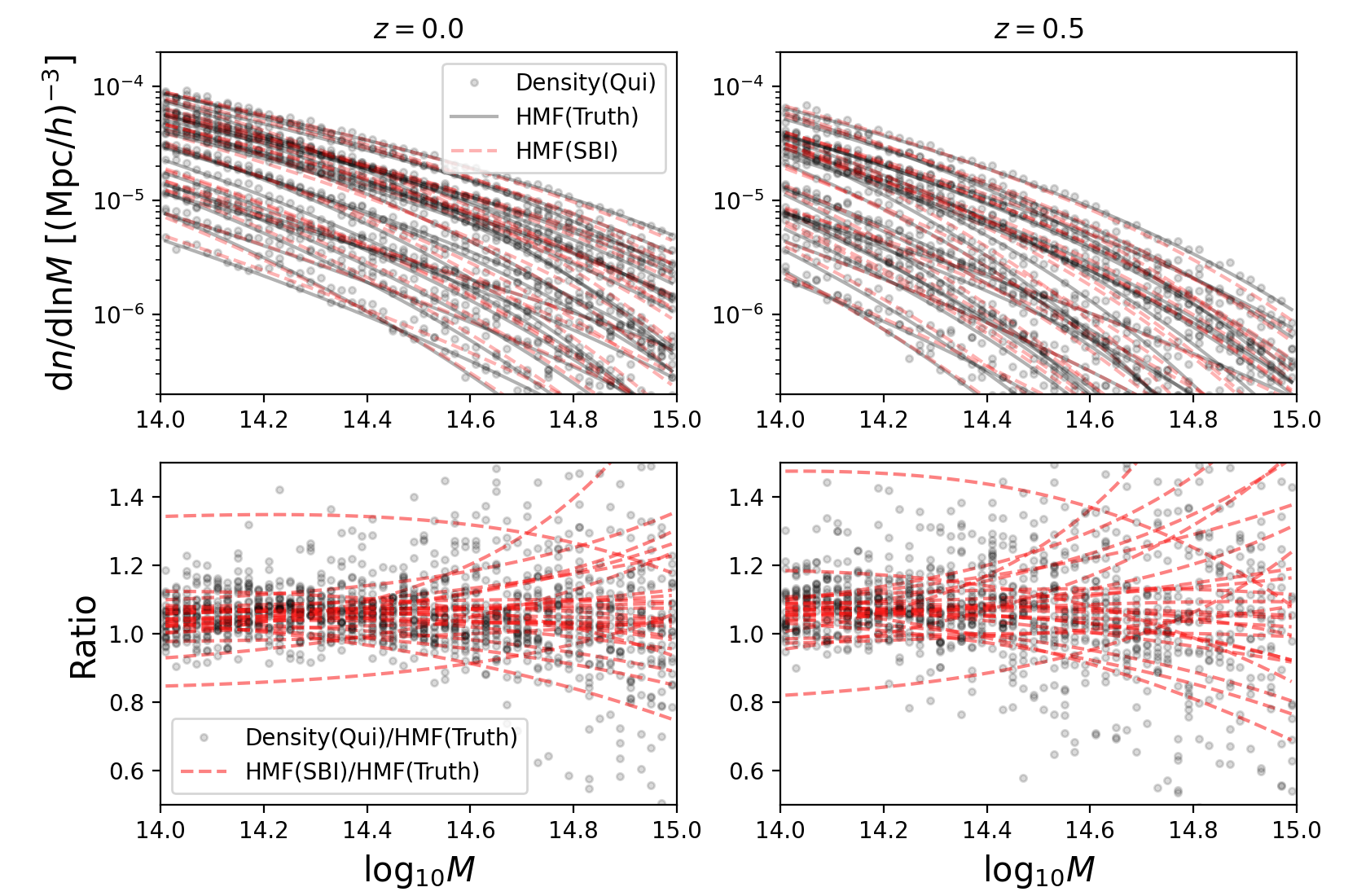}
\caption{
Comparison between HMF and halo mass distribution for the Quijote simulation tests. Upper left panel: Halo mass distributions at redshift 0.0 of 25 sets of Quijote Latin Hypercube simulations (grey data points); HMF computed with the Quijote truth cosmology parameter values (gray solid lines); and  HMF computed with  the mean values of the SBI parameter posterior samples (red dashed lines). Lower left panel: Ratios between the HMFs, and between the HMF and  Quijote halo mass distribution, at redshift 0.0. The HMFs computed with SBI parameter posterior samples better match the Quijote halo mass distributions. Right panels: similar to the left panels but computed at redshift 0.5. See Section~\ref{sec:result_quijote} for more detailed discussions.
}
\label{fig:hmf_sbi}
\end{figure*}

\begin{figure}
\includegraphics[width=\columnwidth]{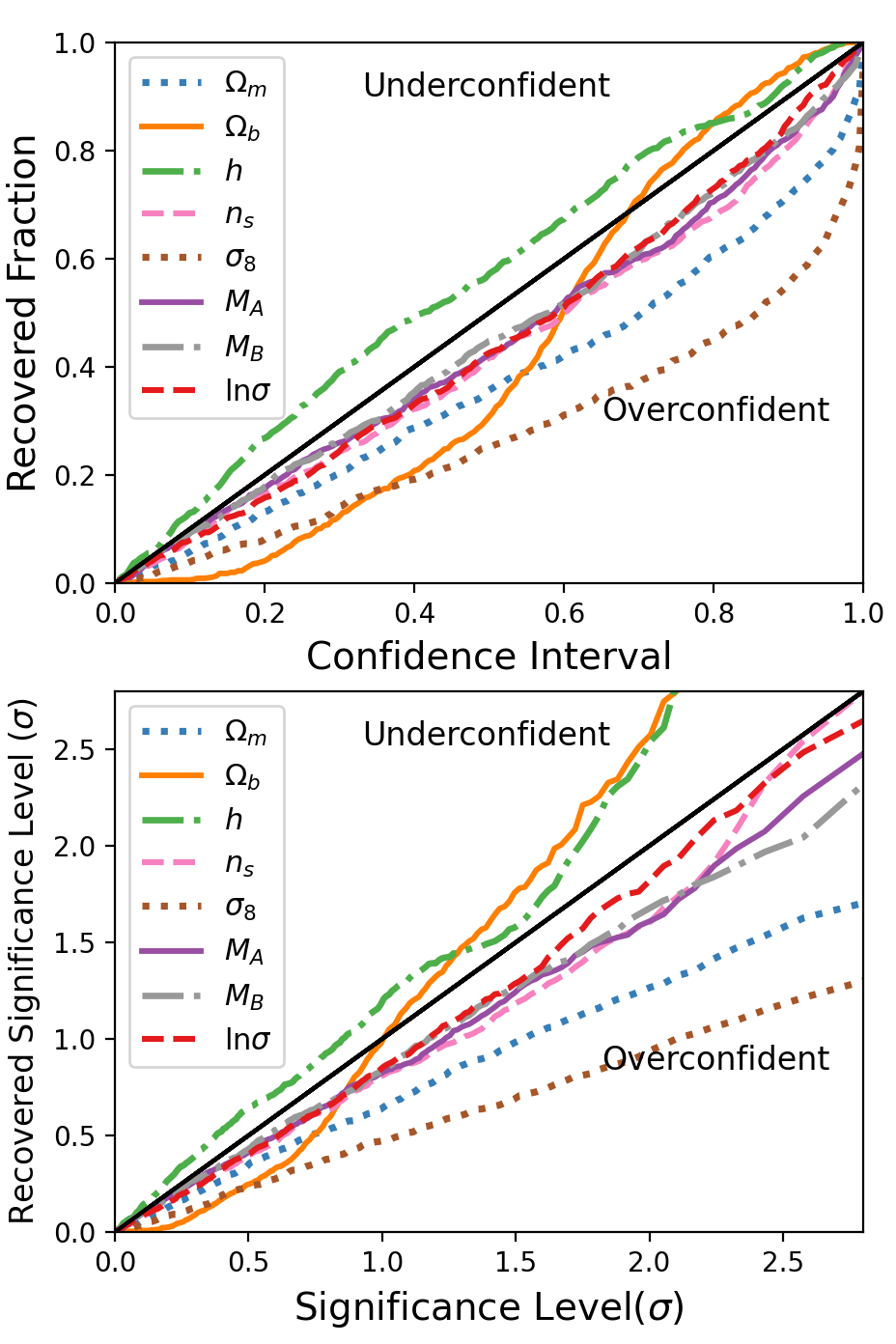}
    \caption{
    Upper panel: Coverage plot showing the fraction of parameter truths recovered on the $y$-axis versus the SBI-estimated confidence interval on the $x$-axis for a model trained on analytical simulations and tested on Quijote simulations. 
    The black solid line represents the ideal scenario. 
    The SBI posterior is significantly overconfident for $\Omega_m$ and $\sigma_8$. 
    The posterior also exhibits some overconfidence for the richness-mass parameters, $M_A$, $M_B$, and $\mathrm{ln}\sigma$. 
    Lower panel: Significance Level Coverage plot.
    The fraction of recovered parameter truth values are converted into a significance level estimation assuming a Gaussian distribution  -- e.g., 2 $\sigma$ significance level indicates 95\% fraction of recovered truth values. Again, SBI behaves overconfidently for $\Omega_m$, $\sigma_8$,  $M_A$, $M_B$, and $\mathrm{ln}\sigma$. See Section~\ref{sec:qui_test} for a more detailed discussion.}
    \label{fig:coverage_ana_qui}
\end{figure}

We apply the methods to test sets from the Quijote simulations.
We consider these Quijote observables to be out-of-domain for both the default SBI and MCMC methods because the halo masses used to derive the observables are from the Quijote N-body simulations, and their constructions are very different from the halo masses sampled from an analytical form of a halo mass function. 
We use this application to demonstrate how SBI and MCMC perform in a situation when the actual observables have a different nature with the training/modeling data. 
If applying SBI to real cluster galaxy abundance observables from our Universe, the real nature of our Universe will be unknown, and thus also out-of-domain. 
We use the test carried out in this subsection to gain insight into how the SBI and MCMC method might behave with real Universe observables.

\subsection{Distribution of Posterior Parameter Samples} 
\label{sec:qui_posterior}

We select three test cases from the Quijote simulations to apply the SBI and MCMC methods to. 
The default SBI and MCMC methods are applied with the same analytical models, training sets and procedures as described in the previous two sections. 
In addition, for comparison with the default SBI method, we also train a separate SBI method, denoted as SBI-Q in this paper, with Quijote-based simulations and apply the SBI-Q method to the Quijote test cases. The training and testing data sets are consistent for the SBI-Q method and thus it will be an in-domain application. 

Figure~\ref{fig:Qui3methods_columns} presents the posteriors from the three methods, default SBI, MCMC, and SBI-Q for three test cases from Quijote simulations. 
For the three methods, the truth values of the parameters are all recovered within the 2$\sigma$ confidence intervals. However, we observe some interesting trends. 
First, for $\Omega_b$, $h$, and $n_s$, we observe similar posterior distributions from the default SBI and SBI-Q method, which are very different from the posterior distributions from the MCMC method. 
As shown in Section~\ref{sec:results}, the default SBI method tends to yield similar posterior distributions for those parameters regardless of their truth values, potentially due to the lack of constraining power from the observables. From this perspective, it is not surprising that the SBI and SBI-Q methods yield similar posterior distributions when the algorithm is insensitive to these parameter's truth values in the training sets. In terms of the MCMC method, the posterior distributions from the MCMC method also share similar trends with the MCMC method applied to analytical tests: the distributions are skewed to one side of the prior range for $\Omega_b$, $h$, and $n_s$. 
For the rest of the parameters, $\Omega_m$, $\sigma_8$, $M_A$, $M_B$, and $\mathrm{log}\sigma$, the SBI methods and the MCMC method all yield informative posterior distributions. 

Interestingly, the SBI-Q method tends to yield posterior distributions that are slightly offset from the default SBI and MCMC method for $\Omega_m$ and $\sigma_8$. These two parameters mainly affect the halo mass function used to derive the observables, and the SBI-Q uses a different set of halo masses from the default SBI and MCMC method. 
Thus, the differences in the posteriors for these two parameters from the three methods are likely driven by the different halo mass functions (see Section~\ref{sec:qui_test}). 
On the other hand, the Quijote simulations, the analytical simulations, and the MCMC method use the same richness-mass relations.
The constraints on the richness-mass relation parameters,  $M_A$, $M_B$, and $\mathrm{log}\sigma$ tend to be more consistent between the three methods. 
These behaviors reflect the importance of physical model selections used in the SBI and MCMC methods.

\subsection{Statistical Testing of SBI Posterior Samples}
\label{sec:qui_test}

In this section, we present the results of SBI applied to Quijote simulations. 
The findings strongly suggest that the default SBI method rained with analytical simulations yields biased results on Quijote-derived observables, with the biases most notable for the $\Omega_m$ and $\sigma_8$ parameters. We discuss these findings below.

Figure~\ref{fig:posterior_1to1_Qui} shows the bias and uncertainty plot that compares the SBI posterior parameter values and their truth values.
In terms of the five informative parameters $\Omega_m$, $\sigma_8$, $M_A$, $M_B$ and $\mathrm{ln}\sigma$, the results appear biased for $\Omega_m$ and $\sigma_8$, where the inferred posterior means are frequently larger than the truth values --- sometimes biased as high as 5$\sigma$. 
Further investigation points to the discrepancies in the HMFs as the source of such large biases. We compute analytical HMFs using the truth parameter values from the Quijote simulations, and compare them to the Quijote halo mass distributions in  Figure~\ref{fig:hmf_sbi}. We find that the Quijote halo mass distributions (gray points) systematically deviates from the analytical HMFs with the truth parameter values. 
However, with the biased cosmology posterior estimations from the SBI, the HMF models (computed with the posterior parameter means) better match the distribution in the Quijote simulations. 

For the $M_A$, $M_B$, and $\mathrm{ln}\sigma$ parameters, there are no obvious signs of bias.
However, in a couple of cases, the truth values are outside the 2$\sigma$ ranges of the posterior distributions, slightly higher than the frequency expected (1 out of 50) for this confidence interval, which indicates that the algorithm may be overly confident or slightly biased on these parameters. 

We further examine the coverage plot of the default SBI application on Quijote observables, as shown in Figure~\ref{fig:coverage_ana_qui}. The algorithm appears to be severely overconfident for the $\sigma_8$ parameter, which is not surprising given the bias we have seen in Figure~\ref{fig:posterior_1to1_Qui}. The algorithm has not been able to recover the truth values of the $\sigma_8$ parameters with sufficiently high frequency in its predicted 1$\sigma$ or 2$\sigma$ ranges. Similarly, we also see the algorithm being severely overconfident for the $\Omega_m$ parameter, for which we have also observed high biases. For the $M_A$, $M_B$ and $\mathrm{ln}\sigma$ parameters, the algorithm behaves better, but is still overconfident with these parameters. 

It is not feasible to run similar studies with the MCMC method on Quijote simulations, given the long time needed to derive MCMC posterior chains. However, given the similarities in MCMC results with these default SBI results, we expect a similar bias in $\Omega_m$ and $\sigma_8$ parameters with the MCMC methods. 
As shown in Figure~\ref{fig:hmf_sbi}, these biases likely stem from mismatches between the analytical HMF (used in the default SBI and MCMC methods) and the Quijote simulations.

\section{Conclusions and Discussions}
\label{sec:conclusiondiscussions}

In this paper, we compare two inference methods, Simulation-Based Inference (SBI) and Markov Chain Monte Carlo (MCMC) applied to parameter inference in optical cluster cosmology.
For training the models, we use an analytical simulator to generate galaxy cluster observables --- number counts and average masses in cluster richness bins.
We test the models on the analytical simulations and on the Quijote Latin Hypercube N-body simulations.
To construct a likelihood function for the MCMC analysis, we use the same analytical models as that used for the analytical simulations to generate predictions for the observable data vector.

For parameters known to be strongly constrained by this observable data vector --- $\Omega_m$, $\sigma_8$, and the cluster richness-mass relation parameters --- we find that SBI can yield posterior parameter samples accurately and efficiently.  
The SBI results are consistent with those from an MCMC analysis. 
This comparison suggests that SBI has the potential to function as an MCMC replacement with accurate posterior sampling for parameters relevant to cluster cosmology.

In contrast, we do not see such consistency in posterior samples for cosmological parameters that we do not expect the observed data vector to constrain --- e.g., $\Omega_b$ and $h$.  In particular, we find that the SBI and MCMC posterior samples differ from one another.  The true parameter values are not accurately captured by the SBI confidence intervals. The behavior of SBI on these parameters likely reflects the inference limitations of the method.

Finally, we also perform out-of-domain tests for the SBI and MCMC methods.  
Here, we perform inference on data vectors derived from Quijote simulations whose halo mass function shapes differ from the analytical forms used in the MCMC application and in generating SBI training sets (i.e. the analytically-derived simulations).
We find that the SBI and MCMC results have comparable posterior samples, with a similar uncertainty and bias.  
Note, it is not feasible to thoroughly test for biases and calibration of posterior sampling with MCMC in the out-of-domain case due to its long-computing time.  
However, we perform such tests with SBI and find that the default SBI method tends to yield biased and overconfident constraints on the $\Omega_m$ and $\sigma_8$ parameters when applied to thousands of test cases from the Quijote simulations. 
The bias appears to be due to the slightly different halo mass functions in the training dataset (based on the analytical simulations) and the testing dataset (based on the Quijote simulations).
SBI produces slightly higher values of $\Omega_m$ and $\sigma_8$. 
Halo mass function predictions from the analytical models with these higher values are more consistent with the halo mass distribution in the Quijote simulations. 
On the other hand, we employ the same richness-mass relation in both the Quijote and analytical simulations of the observed data vector. We do not find any obvious bias in the posterior distribution of the richness-mass relation parameters inferred by the default SBI method. This analysis also illustrates the power of SBI for efficient testing and calibration for a situation where the application of MCMC is less efficient.
In the future, it could be interesting to explore combining training sets generated with different physical models to account for model uncertainties in constructing SBI simulations. 

In general, after an SBI inferrer has been trained with a set of simulations, the posterior distributions from SBI are efficient to compute and easy to test across a wide range of test datasets.  In particular, SBI inference can enable stringent quantification of posterior calibration. The MCMC posterior sampling process takes a much longer time, making it harder to run on similarly large volumes of testing datasets. 
However, the end-to-end SBI method is not necessarily computing-light. The method relies on generating very large sets of simulations that span parameter spaces of interest for training. The simulation generation stage can take a large amount of computing time, particularly as the simulation incorporates more complexity to capture nuances in real data.  In our proof-of-concept, we have limited our simulations to simplified, analytical models that are relatively cheap to produce. 
In addition, the training time required by SBI often increases as the training simulation increases in complexity.  We leave such tests to future work.


\section*{Acknowledgements}

We have made extensive usage of existing software packages, including \texttt{Python}, \texttt{SBI} \citep{tejero2020sbi}, \texttt{PyTorch} \citep{ketkar2021introduction}, \texttt{EMCEE} \citep{2013PASP..125..306F}, ChainConsumer \citep{Hinton2016}, \texttt{CosmoSIS} \citep{2015A&C....12...45Z} \texttt{Astropy} \citep{astropy:2013, astropy:2018, astropy:2022}, \texttt{Numpy} \citep{harris2020array}, \texttt{Scipy} \citep{2020SciPy-NMeth}.

The work of MR is supported by Louis Strigari in the form of graduate research assistantship as well as URA Visiting Scholars Award. This award was used for traveling to Fermi National Accelerator Laboratory which facilitated collaboration with the Deep Skies Lab members. The work of YZ is supported by the Texas A\&M Mitchell institute, and NSF NOIRLab, which is managed by the Association of Universities for Research in Astronomy (AURA) under a cooperative agreement with the U.S. National Science Foundation.  CA acknowledges support from DOE grant DE-SC009193, the Leinweber Center for Theoretical Physics, and the IDEA Scholar Program of the Flatiron Institute.

We acknowledge the Deep Skies Lab as a community of multi-domain experts and collaborators who have facilitated an environment of open discussion, idea generation, and collaboration. This community was important for the development of this project. We thank Dr. Brian Nord for the many helpful discussions, reviews and supervision of this work.  


\vspace{1em}
{\bf Contributions:}
\begin{itemize}
\item MR: Methodology, Software, Validation, Formal analysis, Investigation, Writing - Original Draft, Writing - Review \& Editing, Visualization
\item YZ: Conceptualization, Methodology, Software, Validation, Formal analysis, Investigation, Resources, Data Curation, Writing - Original Draft, Writing - Review \& Editing, Visualization, Supervision
\item CA: Conceptualization, Methodology, Writing - Original Draft, Writing - Review \& Editing
\item LS: Resources, Writing - Review, Supervision, Project administration, Funding acquisition
\item SS: Software, Validation
\item FVN: Methodology, Writing - Review
\end{itemize}

 

\bibliographystyle{mnras}
\bibliography{example} 



\begin{appendix}

\section{SBI Trained with the MAF algorithm}
\label{ap:MAF}

\begin{figure*}
\begin{center}
\includegraphics[width=1.0\columnwidth]{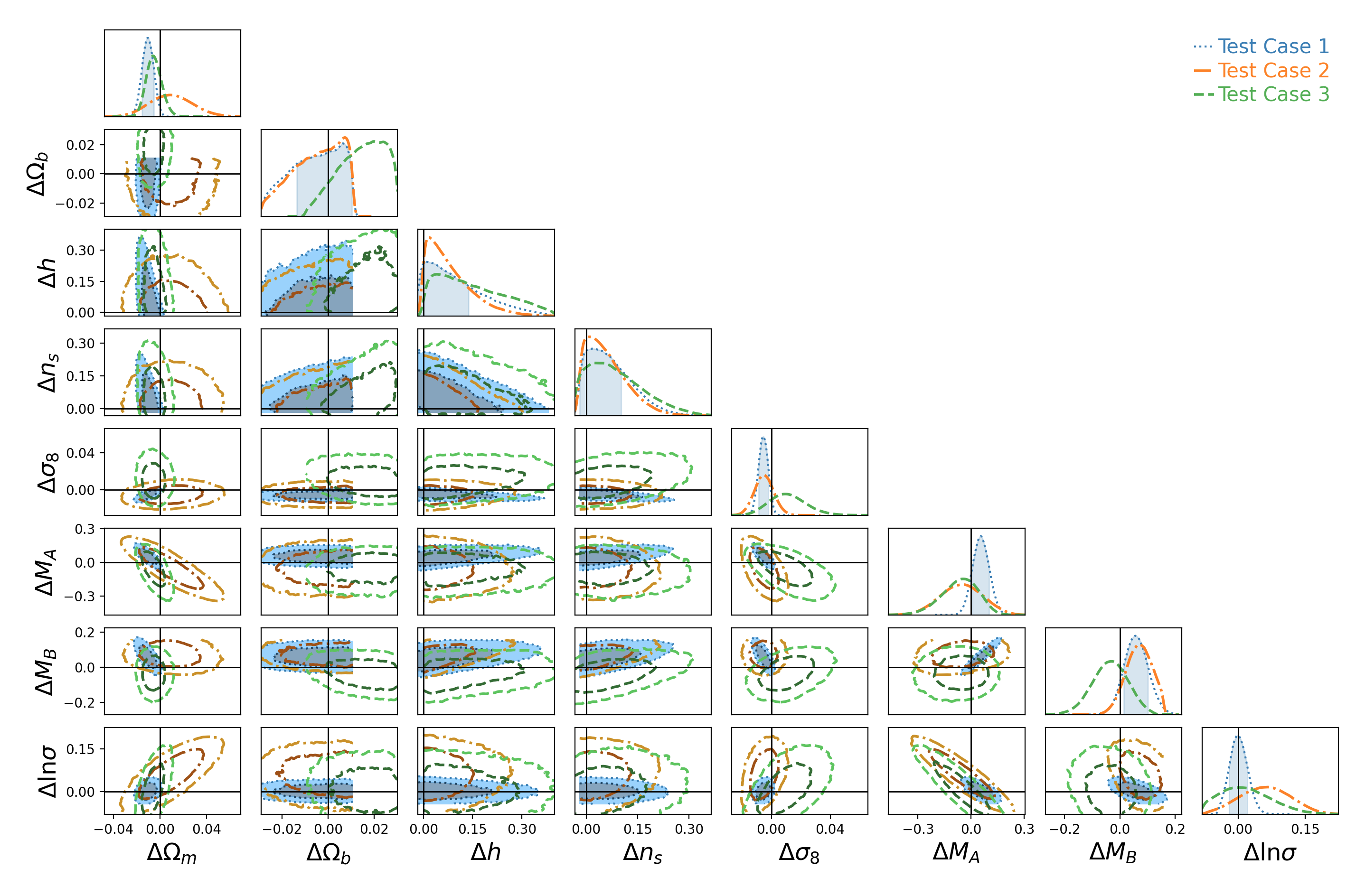}
    \caption{
    Corner plot showing the deviation of SBI parameter posterior samples from the parameter truth values for three test cases (listed in Table~\ref{tbl:AnaTests}) using the MAF density estimator. 
    This plot is analogous to Figure~\ref{fig:contour_ana_ana} which is based on the MDN estimator. Using the MAF, the truth values fall within the $2\sigma$ contour lines for all eight parameters, and the recovered trends are similar to those derived with the MDN.
    }
    \label{fig:contour_MAF}
\end{center}
\end{figure*}

\begin{figure*}
\begin{center}
\includegraphics[width=1.0\columnwidth]{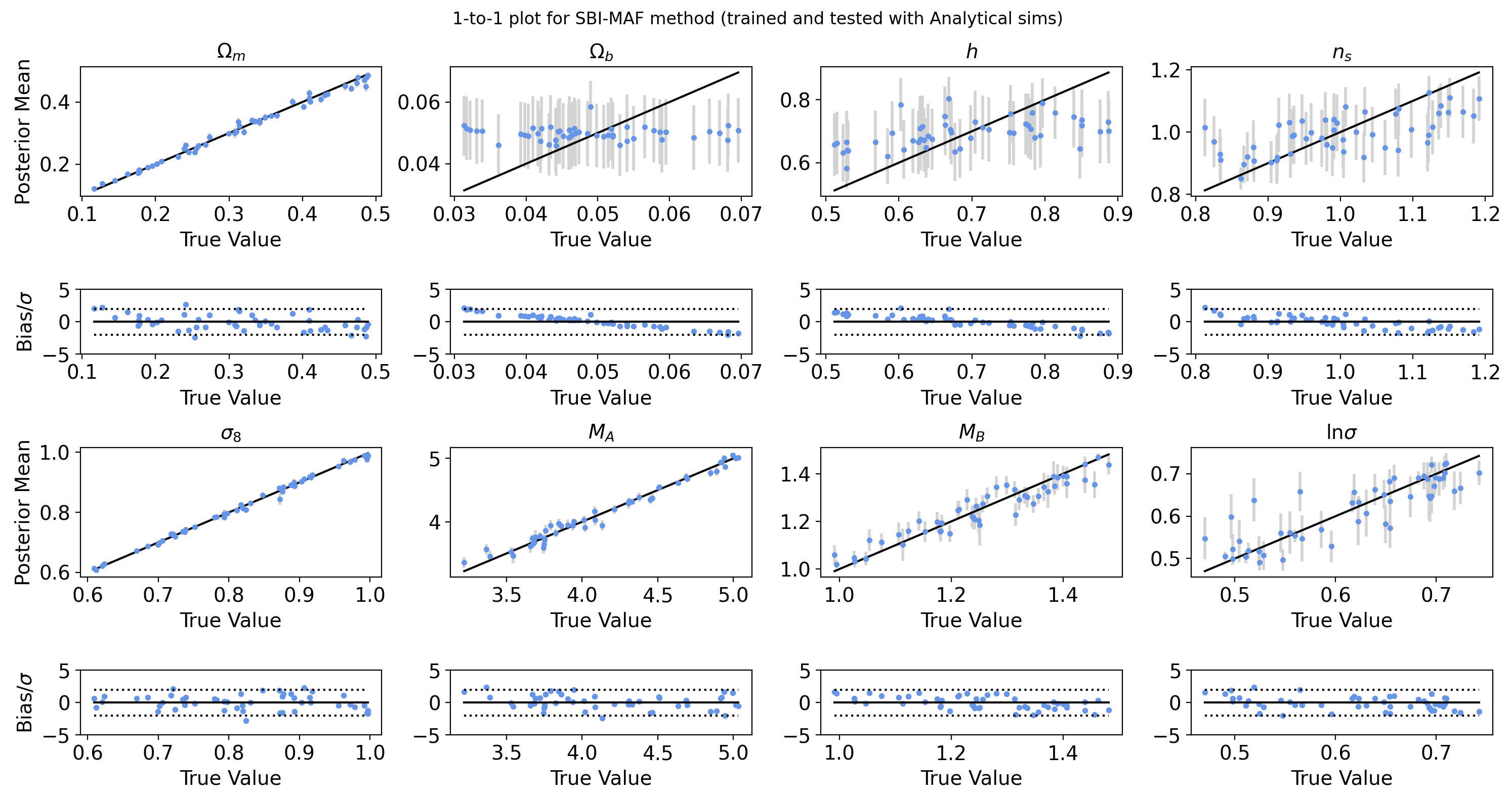}
    \caption{
    This plot shows the mean, bias, and uncertainties of the SBI posterior samples vs the parameter truth values using MAF. This plot is analogous to Figure~\ref{fig:posterior_1to1}, which is based on MDN. 
    Bias and uncertainty levels are similar for MAF and MDN.}
    \label{fig:1to1_MAF}
\end{center}
\end{figure*}

For the primary investigation in this paper, we used Mixture Density Networks (MDNs) to obtain SBI posteriors.
Secondarily, we tested the capacity of the Masked Autoregressive Flow (MAF) on the same task. 
Posterior samples derived with the MAF are shown for three test cases from analytical simulations in Figure~\ref{fig:contour_MAF}, which can be compared to the MDN results in Figure~\ref{fig:contour_ana_ana}.
The posterior distributions acquired from the two density estimators for the cosmological parameters ($\Omega_m$ and $\sigma_8$) and richness-mass relation parameters ($M_A$, $M_B$, and $\mathrm{ln}\sigma$) are similar. 
The marginalized distributions of $\Omega_b$, $h$, and $n_s$ have slightly different shapes.
However, the constraints from both methods recover the truths within the $2\sigma$ uncertainty ranges. 
As discussed in the main body of the paper, the cluster observables do not seem to have strong constraining power for these three parameters, and their posterior distribution shapes likely reflects the methodological inference limits, given our models. 
We also examine the posterior biases and uncertainties of MAF, shown in Figure~\ref{fig:1to1_MAF}. Performance of MAF is very similar to that of the MDN method in this test.

\section{Rank Plots}
\label{sec:app:rank}

\begin{figure*}
\includegraphics[width=1.0\columnwidth]{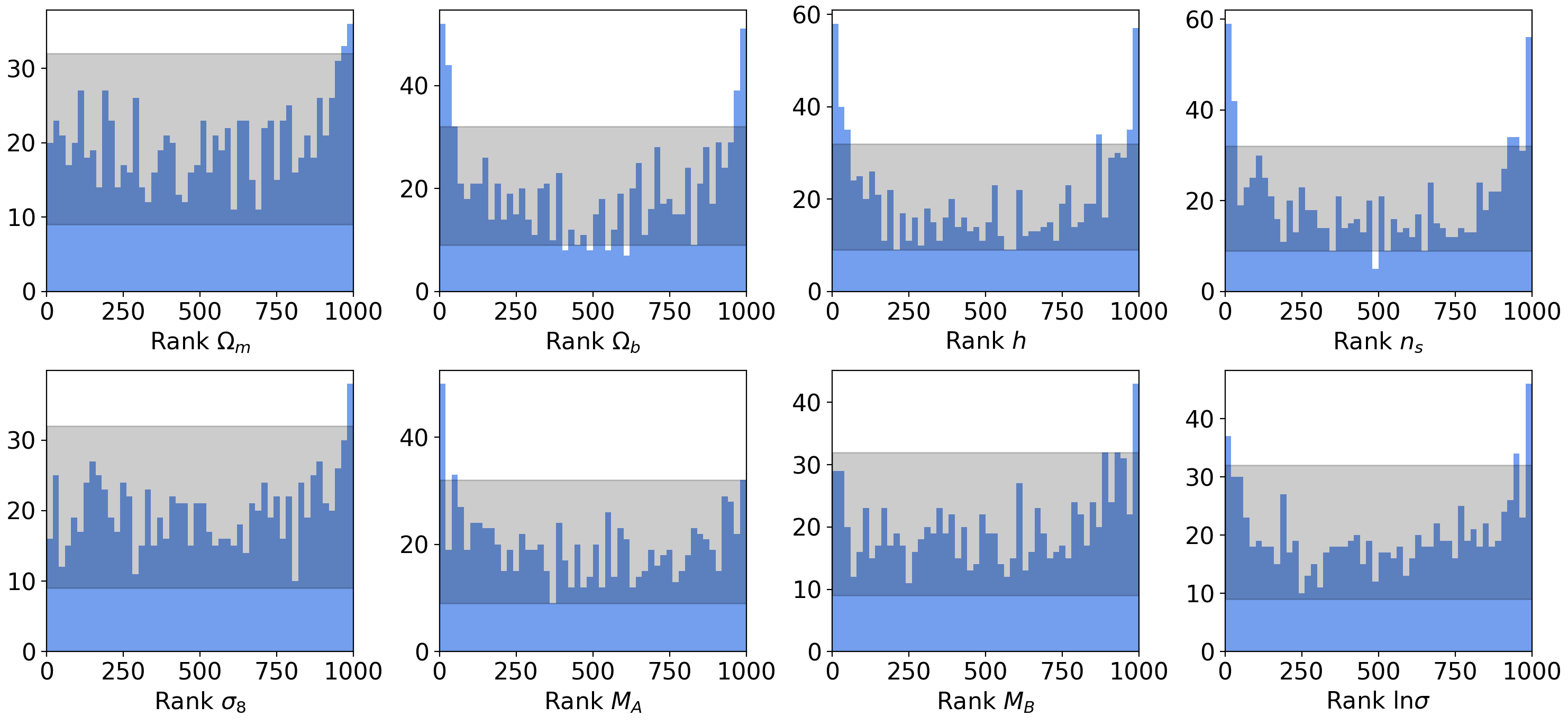}   
    \caption{
    This figure shows the histogram of posterior ranks computed by the default SBI method trained with analytical simulations. 
    The grey shaded region represents the 99\% confidence interval of a uniform rank distribution. We use 1000 tests. 
    For a well-calibrated model, we expect a maximum of 10 tests (1\%) to fall outside the 99\% confidence interval. This condition is satisfied for all eight parameters.
    }
    \label{fig:rank8_ana}
\end{figure*}

\begin{figure*}

\includegraphics[width=1.0\columnwidth]{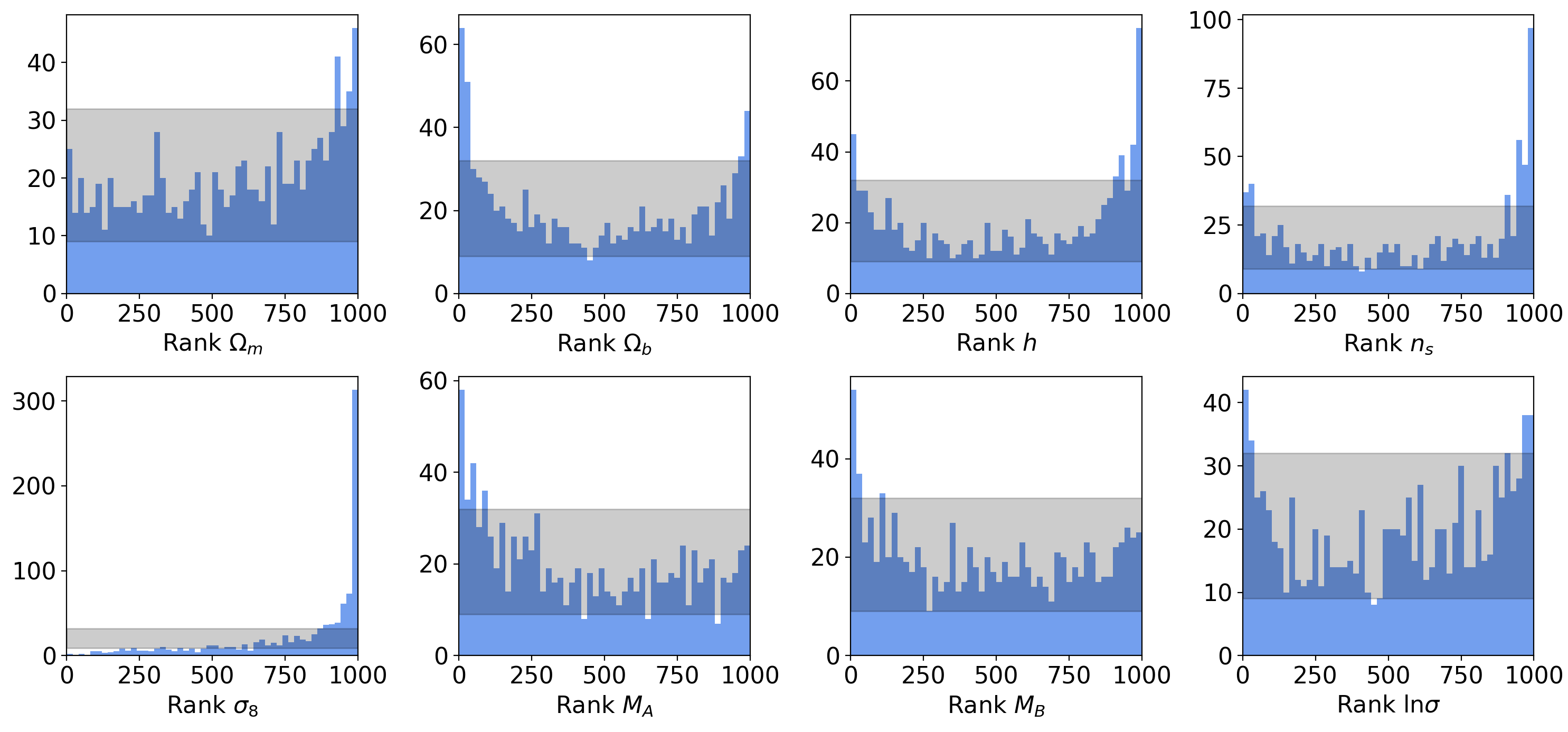}   
\caption{
    This figure shows the histogram of posterior ranks computed for the SBI method trained on analytical simulations and tested on the Quijote simulations.
    The grey shaded region represents the 99\% confidence interval of a uniform rank distribution. 
    We use 1000 tests for this plot. 
    For a well-calibrated model, we expect a maximum of 10 tests (1\%) to fall outside the 99\% confidence interval. 
    The rank distribution of the $\sigma_8$ parameter deviates from this expectation.
}
\label{fig:rank8_Qui}
\end{figure*}

To test the robustness of the SBI posterior samples, we also examine the rank metric~\citep{talts2018validating}, which is another measure of the fidelity of the posterior probability. This rank metric is calculated in the following way: we first acquire 1000 test cases from the analytical simulations or from the Quijote simulations. For each test case, we generate 1000 samples of the parameter values from their posterior distributions. We calculate the rank of each parameter, which refers to the number of posterior parameter sample values that are smaller than the parameter truth value used to generate the simulation. This is done per dimension of the parameters, which is eight for this study. We compile the ranks of all parameters from the 1000 tests. For a properly calibrated posterior, the rank distributions are nearly uniformly distributed. 

We examine ranks by visually examining the histograms. A 99\% confidence interval for the ideal uniform rank histogram is calculated using the binomial distribution.  In Figure \ref{fig:rank8_ana}, we show the rank histogram for the SBI method trained and tested with the analytical simulations. All parameters pass the 99\% confidence interval tests. In Figure \ref{fig:rank8_Qui}, we show the rank distribution for the SBI method trained with the analytical simulations and tested with the Quijote simulations. All parameters, except $\sigma_8$, pass the 99\% confidence interval tests. As revealed by other metrics presented in this paper, the SBI method trained with the analytical simulations and tested on Quijote simulations yields biased results for $\Omega_m$ and $\sigma_8$.

\section{Sensitivity of the SBI result to noises in the observable data vector}
\label{app:sec:sensitivity}

\begin{figure*}
\includegraphics[width=1.0\columnwidth]{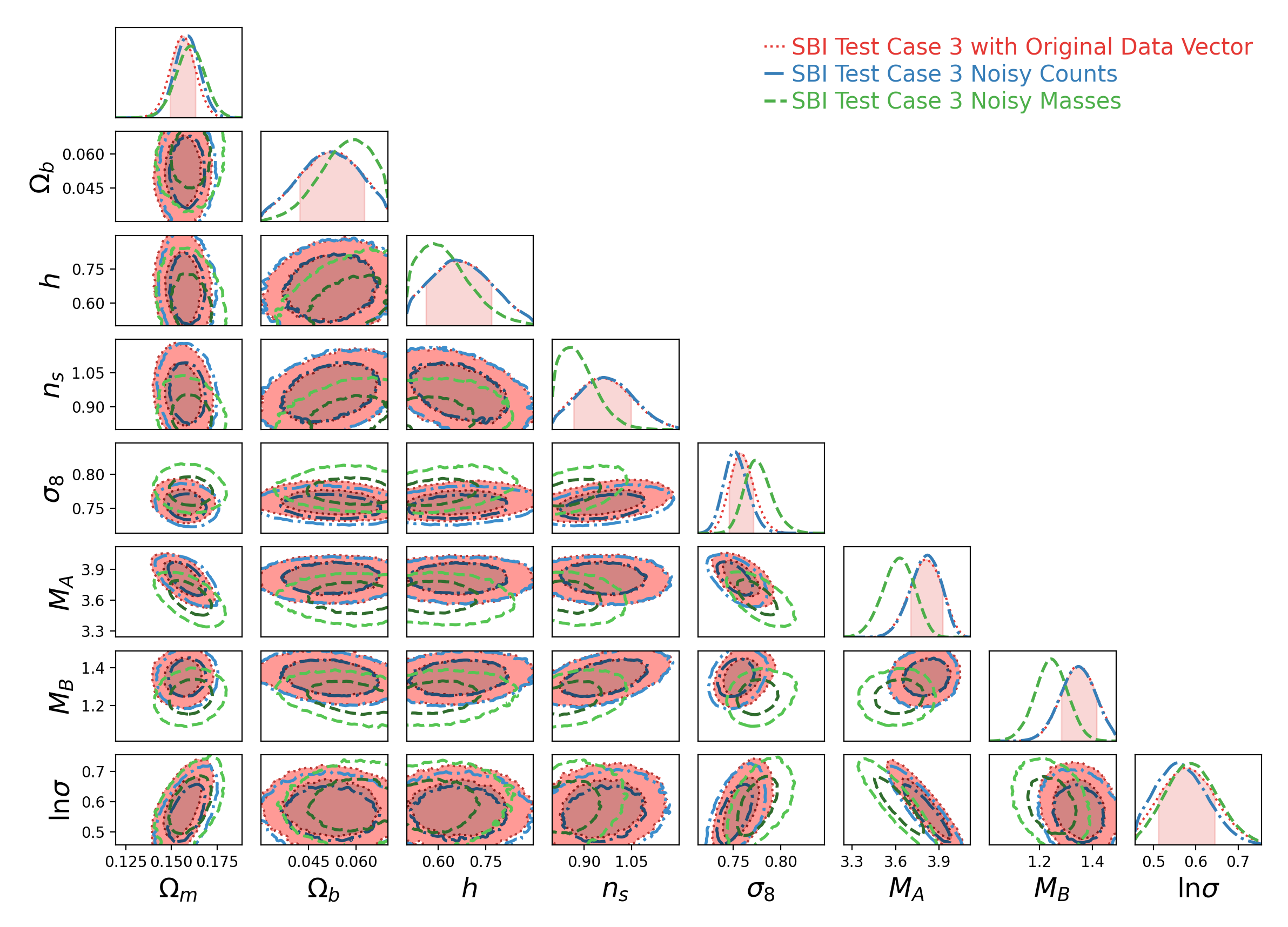}
\caption{
Corner plot showing the posterior samples derived with the default SBI method for Test Case 3 in Table~\ref{tbl:AnaTests} when perturbing the original data vector. 
The red dotted contours show the posterior sample corresponding to  the original data vector, while the blue solid and green dashed contours show the posteriors inferred from the perturbed data vectors.
Perturbing the mass component of the data vector tends to shift  more parameters than perturbing the cluster counts component. 
}
\label{fig:noisyMCMC}
\end{figure*}

In Section~\ref{sec:mcmc_PP}, we show that the posterior predictions for the cluster mass component in the data vector deviate more than the posterior predictions for the cluster count component. 
We speculate that this is because our parameter posterior distributionss are more sensitive to fluctuations in the cluster count component than the mass component, hence the latter is more tolerated by the modeling. 

We test this hypothesis by perturbing the cluster count and the mass components of the test data vector and re-derive the posterior parameter samples.
We add noise to the observable data vector , following a Gaussian distribution of $\mathcal{N}{(0, S(\theta)}$), where $S(\theta)$ is the covariance matrix used in Section~\ref{sec:mcmc_method}) for either the cluster counts or masses. 
We then re-run the posterior inference process with the perturbed data vectors. 
The posterior parameter samples for one test simulation are shown in Figure~\ref{fig:noisyMCMC}. 
We find that perturbing the cluster counts mainly shifts the posterior distributions of $\Omega_m$ and $\sigma_8$, while perturbing the cluster masses tends to shift both the latter parameters and those of the richness-mass relation ($M_A$, $M_B$, and $\mathrm{ln}\sigma$) parameters. 
The involvement of more parameters in modeling the masses may indicate a higher level of flexibility of the models to accomodate mass fluctuations.

\section{Results from SBI method trained on Quijote Simulations}
\label{app:sec:quijotetrain}

\begin{figure*}
\begin{center}
\includegraphics[width=1.0\columnwidth]{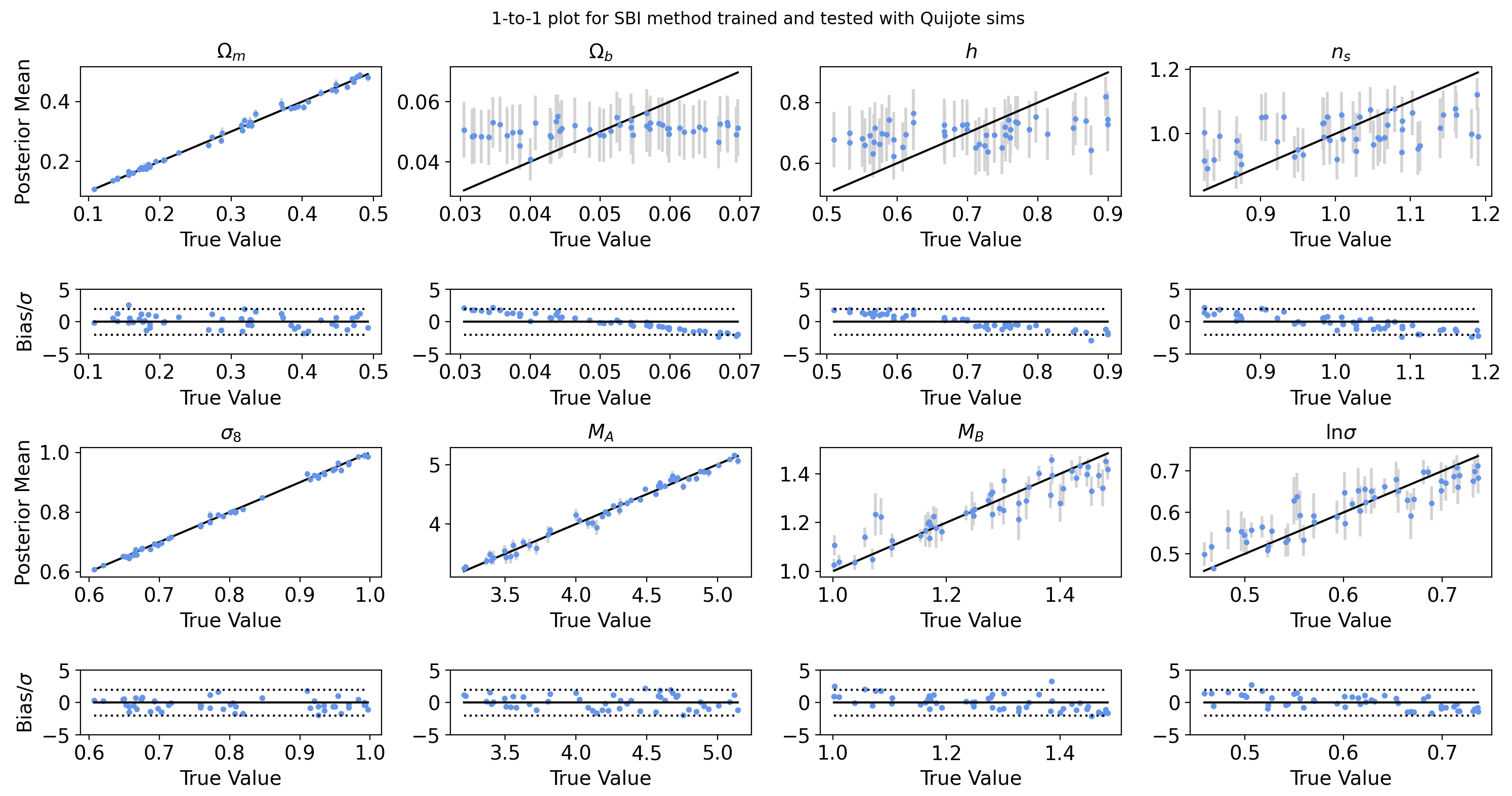}
    \caption{
    This plot shows the mean, bias, and uncertainties of the SBI posterior samples trained and tested with the Quijote simulations. This plot is analogous to Figure~\ref{fig:posterior_1to1}, which is for the default SBI method applied to the analytical simulations. 
    Bias and uncertainty levels are similar in both in-domain applications.
    }
    \label{fig:1to1_QuiTrainTest}
\end{center}
\end{figure*}

In this section, we discuss the biases and uncertainties (Figure~\ref{fig:1to1_QuiTrainTest}) for the SBI-Q model trained and tested on the Quijote simulations. The performance of the SBI-Q model is similar to the default SBI method tested on analytical simulations (Figure~\ref{fig:posterior_1to1}). 
There is no noticeable bias in contrast to the case of the out-of-domain application, where the default SBI model is tested on the Quijote simulations (Figure~\ref{fig:posterior_1to1_Qui}).  

\end{appendix}


\end{document}